\documentclass[aip,graphicx,amsmath, amsthm, amssymb,twocolumn, 10pt]{revtex4-1}
\usepackage{adjustbox}
\usepackage{float,epsfig}
\usepackage{graphicx}
\usepackage{yfonts}
\usepackage{rotating}
\usepackage{subfigure}
\usepackage{appendix}
\usepackage{comment}
\usepackage{enumitem}
\usepackage{color}

\usepackage{float}
\makeatletter
\let\newfloat\newfloat@ltx
\makeatother
\usepackage{algorithm}

\usepackage[utf8]{inputenc}
\usepackage{braket}
\usepackage{qcircuit}
\usepackage[overload]{empheq}
\usepackage{hyperref}
\usepackage{xcolor}
\usepackage[english]{babel}

\usepackage{algcompatible}
\usepackage[noend]{algpseudocode}
\usepackage{mathtools}

\usepackage{xparse}

\NewDocumentCommand{\INTERVALINNARDS}{ m m }{
    #1 {,} #2
}
\NewDocumentCommand{\interval}{ s m >{\SplitArgument{1}{,}}m m o }{
    \IfBooleanTF{#1}{
        \left#2 \INTERVALINNARDS #3 \right#4
    }{
        \IfValueTF{#5}{
            #5{#2} \INTERVALINNARDS #3 #5{#4}
        }{
            #2 \INTERVALINNARDS #3 #4
        }
    }
}

\usepackage{lipsum}

\usepackage{qcircuit}



\makeatletter
\def\@email#1#2{%
 \endgroup
 \patchcmd{\titleblock@produce}
  {\frontmatter@RRAPformat}
  {\frontmatter@RRAPformat{\produce@RRAP{*#1\href{mailto:#2}{#2}}}\frontmatter@RRAPformat}
  {}{}
}%
\makeatother

\begin{document}
\makeatletter 
 \def\@eqnnum{{\normalsize \normalcolor (\theequation)}} 
  \makeatother
  
\newtheorem{theorem}{\bf Theorem}[section]
\newtheorem{proposition}[theorem]{\bf Proposition}
\newtheorem{definition}[theorem]{\bf Definition}
\newtheorem{corollary}[theorem]{\bf Corollary}
\newtheorem{example}[theorem]{\bf Example}
\newtheorem{exam}[theorem]{\bf Example}
\newtheorem{remark}[theorem]{\bf Remark}
\newtheorem{lemma}[theorem]{\bf Lemma}
\newtheorem{statement}[theorem]{\bf Statement}
\newcommand{\nrm}[1]{|\!|\!| {#1} |\!|\!|}

\newcommand{\calL}{{\mathcal L}}
\newcommand{\calX}{{\mathcal X}}
\newcommand{\calA}{{\mathcal A}}
\newcommand{\calB}{{\mathcal B}}
\newcommand{\calC}{{\mathcal C}}
\newcommand{\calK}{{\mathcal K}}
\newcommand{\C}{{\mathbb C}}
\newcommand{\R}{{\mathbb R}}
\newcommand{\U}{{\mathrm U}}
\renewcommand{\SS}{{\mathbb S}}
\newcommand{\LL}{{\mathbb L}}
\newcommand{\st}{{\star}}
\def\kernel{\mathop{\rm kernel}\nolimits}
\def\sigan{\mathop{\rm span}\nolimits}

\newcommand{\klasse}{{\boldsymbol \Delta}}

\newcommand{\ba}{\begin{array}}
\newcommand{\ea}{\end{array}}
\newcommand{\von}{\vskip 1ex}
\newcommand{\vone}{\vskip 2ex}
\newcommand{\vtwo}{\vskip 4ex}
\newcommand{\dm}[1]{ {\displaystyle{#1} } }

\newcommand{\be}{\begin{equation}}
\newcommand{\ee}{\end{equation}}
\newcommand{\beano}{\begin{eqnarray*}}
\newcommand{\eeano}{\end{eqnarray*}}
\newcommand{\inp}[2]{\langle {#1} ,\,{#2} \rangle}
\def\bmatrix#1{\left[ \begin{matrix} #1 \end{matrix} \right]}
\def \noin{\noindent}
\newcommand{\evenindex}{\Pi_e}

\newcommand{\tb}[1]{\textcolor{blue}{ #1}}
\newcommand{\tm}[1]{\textcolor{magenta}{ #1}}
\newcommand{\tre}[1]{\textcolor{red}{ #1}}
\newcommand{\snote}[1]{\textcolor{blue}{Shantanav: #1}}



\def \K{{\mathbf k}}
\def \N{{\mathbb N}}
\def \R{{\mathbb R}}
\def \F{{\mathbb F}}
\def \C{{\mathbb C}}
\def \Q{{\mathbb Q}}
\def \Z{{\mathbb Z}}
\def \I{{\mathbb I}}
\def \D{{\mathcal D}}
\def \H{{\mathcal H}}
\def \P{{\mathcal P}}
\def \M{{\mathcal M}}
\def \B{{\mathcal B}}
\def \O{{\mathcal O}}
\def \calG{{\mathcal G}}
\def \PO{{\mathcal {PO}}}
\def \X{{\mathcal X}}
\def \Y{{\mathcal Y}}
\def \calW{{\mathcal W}}
\def \pf{{\bf Proof: }}
\def \lam{{\lambda}}
\def\lc{\left\lceil}   
\def\rc{\right\rceil}
\def \N{{\mathbb N}}
\def \Ls{{\Lambda}_{m-1}}
\def \Gb{\mathrm{G}}
\def \Hb{\mathrm{H}}
\def \Delta{\triangle}
\def \Rar{\Rightarrow}
\def \p{{\mathsf{p}(\lam; v)}}

\def \D{{\mathbb D}}

\def \tr{\mathrm{Tr}}
\def \cond{\mathrm{cond}}
\def \lam{\lambda}
\def \sig{\sigma}
\def \sign{\mathrm{sign}}

\def \ep{\epsilon}
\def \diag{\mathrm{diag}}
\def \rev{\mathrm{rev}}
\def \vec{\mathrm{vec}}

\def \ham{\mathsf{Ham}}
\def \herm{\mathsf{Herm}}
\def \sym{\mathsf{sym}}
\def \odd{\mathsf{sym}}
\def \en{\mathrm{even}}
\def \rank{\mathrm{rank}}
\def \pf{{\bf Proof: }}
\def \dist{\mathrm{dist}}
\def \rar{\rightarrow}

\def \rank{\mathrm{rank}}
\def \pf{{\bf Proof: }}
\def \dist{\mathrm{dist}}
\def \Re{\mathsf{Re}}
\def \Im{\mathsf{Im}}
\def \re{\mathsf{re}}
\def \im{\mathsf{im}}

\def \sym{\mathsf{sym}}
\def \sksym{\mathsf{skew\mbox{-}sym}}
\def \odd{\mathrm{odd}}
\def \even{\mathrm{even}}
\def \herm{\mathsf{Herm}}
\def \skherm{\mathsf{skew\mbox{-}Herm}}
\def \str{\mathrm{ Struct}}
\def \eproof{$\blacksquare$}

\def \cnot{\mathrm{CNOT}}

\def \bS{{\bf S}}
\def \cA{{\cal A}}
\def \E{{\mathcal E}}
\def \X{{\mathcal X}}
\def \F{{\mathcal F}}
\def \cH{\mathcal{H}}
\def \cJ{\mathcal{J}}
\def \tr{\mathrm{Tr}}
\def \range{\mathrm{Range}}
\def \adj{\star}

\pdfstringdefDisableCommands{%
  \def\\{}%
  \def\texttt#1{<#1>}%
}
\def \adj{\star}

\def \pal{\mathrm{palindromic}}
\def \palpen{\mathrm{palindromic~~ pencil}}
\def \palpoly{\mathrm{palindromic~~ polynomial}}
\def \odd{\mathrm{odd}}
\def \even{\mathrm{even}}

\newcommand{\tg}[1]{\textcolor{green}{ #1}}

\preprint{AIP/123-QED}

\title{Scalable quantum circuits for exponential of Pauli strings \\ and Hamiltonian simulations}

\author{Rohit Sarma Sarkar}
\affiliation{Department of Mathematics, Indian Institute of Technology Kharagpur, Kharagpur 721302, India}
\email{ rohit15sarkar@yahoo.com}
\author{Sabyasachi Chakraborty}
\affiliation{Department of Physics, Indian Institute of Technology Kharagpur, Kharagpur 721302, India}
\email{ sabyasachi.sch@gmail.com}
\author{Bibhas Adhikari}
\affiliation{Fujitsu Research of America, Inc.,
Santa Clara, CA, USA}
\email{badhikari@fujitsu.com}




\begin{abstract}
In this paper, we design quantum circuits for the exponential of scaled $n$-qubit Pauli strings using single-qubit rotation gates, Hadamard gate, and CNOT gates. A key result we derive is that any two Pauli-string operators composed of identity and $X$ gates are permutation similar, and the corresponding permutation matrices are product of CNOT gates, with the $n$-th qubit serving as the control qubit. Consequently, we demonstrate that the proposed circuit model for exponential of any Pauli-string is implementable on low-connected quantum hardware and scalable i.e. quantum circuits for $(n+1)$-qubit systems can be constructed from $n$-qubit circuits by adding additional quantum gates and the extra qubit. We then apply these circuit models to approximate unitary evolution for several classes of Hamiltonians using the Suzuki-Trotter approximation. These Hamiltonians include $2$-sparse block-diagonal Hamiltonians, Ising Hamiltonians, and both time-independent and time-dependent Random Field Heisenberg Hamiltonians and Transverse Magnetic Random Quantum Ising Hamiltonians. Simulations for systems of up to 18 qubits show that the circuit approximation closely matches the exact evolution, with errors comparable to the numerical Trotterization error. Finally, we consider noise models in quantum circuit simulations to account for gate implementation errors in NISQ computers, and observe that the noisy simulation closely resembles the noiseless one when gate and idle errors are on the order of $O(10^{-3})$ or smaller.

\end{abstract}

\maketitle
\thispagestyle{empty}


\section{Introduction}\label{introduction}

Simulating the Hamiltonians of many-body systems is a highly anticipated achievement of quantum computers, first envisioned independently by Paul Benioff \cite{benioff1980computer} and Yuri Manin \cite{manin1980computable}, and later popularized by Richard Feynman \cite{feynman2018simulating}. Several methods are proposed to encode the loss functions of optimization problems in the Hamiltonians of quantum systems to obtain near-optimal solutions more efficiently than classical methods \cite{bharti2022noisy}. The quantum circuit model holds great promise for the Hamiltonian simulation of many-body systems using digital quantum computers in the near term \cite{fauseweh2024quantum}. Developing scalable quantum circuit models and methods to implement the unitary evolution defined by Hamiltonians remains one of the primary research areas in quantum computing.

Since Pauli-strings defined by $\sigma_1 \otimes \sigma_2 \otimes \cdots \otimes \sigma_n,$ $\sigma_k\in\{I_2,X,Y,Z\},$ $k=1,\hdots,n$ form an orthonormal basis of the algebra of matrices of order $2^n$, the Hamiltonian operators corresponding to $n$-qubit systems are written as linear combinations of Pauli-string matrices. Here, $X,Y,Z$ are Pauli spin matrices of order $2$ given by \begin{eqnarray*}
    X=\bmatrix{0&1\\1&0}, Y=\bmatrix{0&-\iota\\\iota&0}, Z=\bmatrix{1&0\\0&-1},
\end{eqnarray*} $I_2$ is the identity matrix of order $2$, $\iota=\sqrt{-1},$ and $\otimes$ denotes the Kronecker or tensor product of matrices. From a physics perspective, the  local interactions among the qubits or sub-systems are represented mathematically by the Pauli-strings for Hamiltonians of physical $n$-qubit systems such as non-equilibrium phases of matter \cite{klein2006simulating,hofstetter2018quantum}, strongly correlated systems \cite{peotta2021determination}, chemical reaction mechanisms \cite{mcardle2020quantum,lanyon2010towards}. Thus the total Hamiltonian, which represents the total energy of the system is defined by an operator  $H(t)=\sum_{j=1}^{L} a_j(t)H_j$ $t\in [0,\, T]$, where $a_j:[0, \, T]\rightarrow \R$ is a function of time parameter $t$ and $H_j$s denote Pauli-strings, which are also called  \textit{local Hamiltonians}. If $a_j$ is constant function for all $j$ then the Hamiltonian is called time-independent, otherwise it is called time-dependent.

The evolution of the $n$-qubit system is governed by the Schrödinger equation $\hbar\, \ket{\dot{\psi}(t)}=-\iota  H(t)\ket{\psi(t)},$ where $t\in [0, \, T],$ $T > 0$ is the total evolution time and $\hbar$ is Planck's constant. Then the Hamiltonian simulation is concerned with approximating the quantum states $\ket{\psi(t)}=U(t)\ket{\psi(0)},$ where $U(t)=\exp(-\iota t H)$ (assuming $t$  to be a unit of $\hbar$) for time-independent case and $U(t) = \mathcal{T} \; \exp{(-\iota \int_{0}^{t}H({{t}'})d{{{t}'}})}$ for time-dependent Hamiltonian operators, where  $\mathcal{T}$ denotes a time-ordering \cite{poulin2011quantum}, and $\ket{\psi(0)}$ denotes the initial state of the system.

Since classical algorithms are intractable for performing Hamiltonian simulations for large multipartite physical systems, various quantum approaches are proposed with innovative tools for analyzing simulation errors. For instance, quantum algorithms that have polynomial complexity in the number of qubits have been a focal point in the literature corresponding to specific quantum data structure models \cite{low2018hamiltonian,low2017optimal,haah2021quantum,Dong2022}. Whereas, several series approximation of $U(t)$ including truncated Taylor series \cite{berry2015simulating,Zhong2022}, truncated Dyson series \cite{kieferova2019simulating}; randomized evolution \cite{childs2019faster,faehrmann2022randomizing}; perturbative quantum simulation \cite{sun2022perturbative}; variational techniques \cite{yuan2019theory,yao2021adaptive,benedetti2021hardware,wang2021noise,stilck2021limitations,mc2023classically}; and quantum approximate optimization algorithm (QAOA) \cite{Lotshaw2023} are some of the well-known methods explored in the literature for numerical simulation of Hamiltonians for various physical systems. 

On the other hand, quantum circuit simulation of Hamiltonians is concerned with a quantum circuit approximation of the unitary evolution operator $U(t)$, which is defined above. In this context, $n$-qubit quantum circuit representation of exponential of scaling of local Hamiltonians plays a key role in order to implement $U(t)$ through a quantum circuit by employing a Suzuki-Trotter product type formulae for approximation of $U(t).$ Then with a given initial state $\ket{\psi(0)},$ the states of the system $\ket{\psi(t)}$ can be approximated by quantum circuit output as $U(t)\ket{\psi(0)}$ \cite{haah2021quantum,Dong2022}. Even though the Trotterization techniques have poor error scaling \cite{PhysRevX.11.011020}, tighter error bounds exist for specific systems \cite{PhysRevX.11.011020,childs2019nearly}.  Numerical analyses have shown that these product formulae require significantly fewer entangling gates and $T$ gates compared a few other improved techniques as discussed in the Refs. \cite{PhysRevX.11.011020,childs2018toward}. In another direction, parametrized quantum circuit models are proposed for learning algorithms of parametrized Hamiltonians \cite{wiebe2014hamiltonian,shi2022parameterized}. 

A major aspect of circuit simulation is \textit{scalability}. Here, following \cite{SarmaSarkar2023}, we call the construct of a quantum circuit implementation of the unitary $U(t)$ for an $n$-qubit system is \textit{scalable}, if the corresponding circuit for the $(n+1)$-qubit system can be formed with the addition of some new elementary gates along with an additional qubit to the existing circuit on $n$-qubits without altering it. For quantum circuit simulation on NISQ computers, two additional challenges are: \textit{low-connectivity} and \textit{noise}. The low-connectivity of existing quantum hardware prevents the implementation of controlled-gates in a circuit for arbitrary pair of qubits in the circuit and gives rise to \textit{qubit mapping problem}  \cite{li2019tackling}. Further, existing quantum technologies suffer from short qubit decoherence times and high gate error rates, which result in the presence of random noise during circuit simulations \cite{bharti2022noisy}. Thus, it is important to analyze performance of circuit simulation in presence of noise by considering various noise models to determine the limits of noise that can be allowed to estimate the simulation results with a given tolerance of error.    

There are several proposals for quantum circuit simulation of various Hamiltonians in the literature \cite{Raeisi2012,vanden2020,Clinton2021,Mukhopadhyay2023,Peng2022,Yordanov2020}. Indeed, most of them lack scalability and require fully connected quantum hardware for their implementation. Besides, they lack any analysis on the effects of noise in their proposed quantum circuit simulation. In this paper, we overcome these challenges by developing scalable quantum circuits for $\exp(\iota \theta \sigma)$ with CNOT, Hadamard, and one-qubit rotation gates for any Pauli-string $\sigma$ for $n$-qubit systems  and $\theta\in \R,$ that can be implemented in low-connected quantum Hardware. Then we apply these circuits for quantum circuit simulation of several classes of Hamiltonians using the Suzuki-Trotter approximation of $U(t).$ 



Indeed, let us denote the set of $2^n$ Pauli-strings for $n$-qubit systems formed by $I_2$ and $X$ as $\mathcal{S}_{I,X}^{(n)}$. Then we show that quantum circuit representations of $\exp(\iota \theta (I_2^{\otimes n-1} \otimes X))$ and  permutation matrices for $n$-qubit systems of a specific pattern enable the design of quantum circuits for $\exp(\iota\theta\sigma)$ for any Pauli-string $\sigma.$ Indeed, we show that  $\exp(\iota\theta\tau)=P\exp(\iota \theta (I_2^{\otimes n-1}\otimes X)) P^\dagger$ for any $\tau\in \mathcal{S}_{I,X}^{(n)}$ where $P$ is a permutation matrix which corresponds to the product of a sequence of two-qubit controlled  gates given by $\cnot_{n,j}$ $j<n$, which denotes a CNOT gate with $j$-th qubit as target and $n$-th as the control qubit. Next we show that,  for any Pauli-string $\sigma$,  $\exp(\iota\theta\sigma)=Q\exp(\iota\theta\tau)Q^\dagger$ for some $\tau\in \mathcal{S}_{I,X}^{(n)}$, and $Q$ a tensor product of $I_2,$ $S$ or $H$ gates where $$H=\frac{1}{\sqrt{2}}\bmatrix{1&1\\1&-1},\,\, \mbox{and}\,\, S=\bmatrix{e^{\iota\pi/4} &0 \\ 0 & e^{-\iota\pi/4}}$$ is the rotation gate of a qubit around $Z$-axis by an angle $\pi/4,$ and $\dagger$ denotes conjugate transpose of a matrix. 

It is evident that our proposal of quantum circuit representation of $\exp(i\theta\sigma)$, $\theta\in \R$ for any Pauli-string $\sigma$ for $n$-qubit systems is fundamentally different from the existing techniques  \cite{Raeisi2012,Reiher2017,vanden2020,Yordanov2020,Clinton2021,Kawase2022,Mukhopadhyay2023,Mansky2023}. To be specific, permutation similarity of Pauli-strings formed by $I_2$ and $X$ gates leads to the use of CNOT gates mainly with respect to the control or target qubit as the $n$-th (or last qubit in the $n$-qubit register) in the quantum circuit model along with elementary single qubit rotation gates and Hadamard gate. This is a significant quantum technological advantage since quantum hardware with star graph architecture can facilitate direct implementation of the proposed quantum circuits for exponential of Pauli-strings. The hardware such as IBM Q5 Yorktown/ IBM QX2 for $5$ qubits provide such a setup, where the qubit in the center of the architecture can act as the $n$-th qubit \cite{Ibm2017}.

Then we employ the Suzuki-Trotter product approximation formula of different orders to perform circuit simulation of several classes of 1D Hamiltonian operators that include: $2$-sparse block-diagonal Hamiltonian, Quantum Ising model Hamiltonian, Heisenberg Hamiltonian with random field \cite{pires2021theoretical}, and Transverse magnetic field quantum Ising model with an external random field \cite{pires2021theoretical}. The quantum circuit approximation of the unitary evolution are executed in Qiskit for up to $18$-qubit systems for $2$-sparse block-diagonal Hamiltonian and Quantum Ising model Hamiltonian, and up to $14$-qubit systems for the other Hamiltonians.  

Further, we consider different noise models to investigate the performance of the proposed circuit simulation method for Hamiltonians when the gates used in the circuits are exposed to noise. We compare the fidelity of outputs obtained by noisy and noiseless simulation for time-independent random Field Heisenberg Hamiltonian and transverse magnetic random quantum Ising Hamiltonian for different input states considering ranges of gate error, which includes bit-flip and phase-flip errors, and idle error, which includes amplitude-damping and phase damping error. We observe that the gate error and idle error are to be kept below $O(10^{-3})$ in order to achieve high-fidelity values between the outputs of the noisy quantum circuit and its noiseless counterpart.

The rest of the paper is organized as follows. Section \ref{ckt_model_Pauli_exp} is devoted to quantum circuit design of $\exp(i\theta\sigma)$ for any Pauli-string $\sigma$ of $n$-qubit systems, $\theta\in\R$. Application of these quantum circuits for Hamiltonian simulations employing the Suzuki-Trotter approximation formula is given in Section \ref{hsim_qcktall}. The Section \ref{Noisy_simulation} deals with circuit simulations of time-independent transverse magnetic random quantum Ising Hamiltonian and Random Field Heisenberg Hamiltonian in the existence of noise with different noise models. Finally, a summary of our findings, along with future perspectives are discussed in Section \ref{Conclusion}.Appendix \ref{appdx:A2} includes an example to describe the permutation matrices discussed in Section \ref{ckt_model_Pauli_exp}. Appendix \ref{qc_hamil}, \ref{Noise_Model_details} delineate Suzuki-Trotter expansion and the details of noise models respectively.

\section{Quantum circuit model for scaled exponential of Pauli strings}
\label{ckt_model_Pauli_exp}

In this section, we develop scalable quantum circuit for exponential of Pauli-strings corresponding to $n$-qubit systems. Thus we lay the ground work for Hamiltonian simulation via circuit model of computation employing a Suzuki-Trotter like approximation formula for the exponential of Hamiltonian operators.

First, we introduce the following notations. For a given $x\in\{0,\hdots,2^{n-1}-1\}$ with its binary representation $(x_{n-2},\hdots,x_0)$ such that $x=\sum_{j=0}^{n-2} x_j2^{j}, x_{j}\in\{0,1\} $, we define a circuit in the following way. For each $x_j,$ $0\leq j\leq n-2$ the circuit contains a $\cnot_{(n,n-j-1)}$ gate if the $x_j=1,$ where $\cnot_{(n,n-j-1)}$ denotes a $\cnot$ gate with $n$-th qubit as the control and $(n-j-1)$-th qubit as target. Since any $\cnot$ gate represents a permutation matrix, we define new permutations as product of $\cnot$ gates as follows, which we play a primary role in the sequel.  
We denote these permutations as \begin{eqnarray}\label{permus1}
    \Pi \mathsf{T}^{e}_{n,x}&=&\prod_{j=0}^{n-2}  (\cnot_{(n,n-j-1)})^{\delta_{1,x_j} } \\
    \Pi \mathsf{T}^o_{n,x}&=&(\cnot_{(n-m-1,n)})(\Pi \mathsf{T}^e_{n,x})(\cnot_{(n-m-1,n)}) \nonumber
\end{eqnarray} {where $m$ is the greatest integer $0\leq m\leq n-2$ such that $x_{m}=1$ in the binary string of $x=(x_{n-2}\hdots x_0)$ i.e. $m=\mbox{max}\{j | \delta_{1,x_j}=1\}$ and $\delta$ denotes Kronecker delta function and $(\cnot_{(n,n-j-1)})^{0}$ is considered to be the Identity matrix. For $x=0$, we consider $\Pi \mathsf{T}^e_{n,x}$ and $\Pi \mathsf{T}^e_{n,x}$ as the identity matrix i.e. absence of any $\cnot$ gates.

Then we recall some permutation matrices associated with CNOT gates from \cite{SarmaSarkar2023} \cite{sarkar2024quantum}. The set of binary strings $\{(x_{n-2},x_2,\hdots,x_{0}) : x_j\in\{0,1\}\}$ and the set of all subsets of $[n-1] :=\{1,\hdots,n-1\}$ have a one-one correspondence defined by $\chi: \{0,1\}^{n-1}\rightarrow 2^{[n-1]},$ which assigns $x=(x_{n-2},x_{n-3},\hdots,x_{0})$ to $\chi(x)=\Lambda_x :=\{j: x_j=1, 1\leq j\leq n-1\}\subseteq [n-1].$ Thus each position of the string represents a characteristic function for $\Lambda_x.$ Then we rewrite the following theorem from \cite{sarkar2024quantum}.

\begin{theorem}\label{constructeven}
Let $\chi: \{0,1\}^{n-1}\rightarrow 2^{[n-1]}$ be the bijective function as defined above such that $\chi(x)=\Lambda_x$. For any $x\equiv (x_{n-2},\hdots,x_0)\in\{0,1\}^{n-1},$ define the functions $\alpha_{\Lambda_x}^g:  \{0,1\}^{n-1}\rightarrow \{0,\hdots,2^{n-1}-1\}$ and  $\beta_{\Lambda_x}^g:  \{0,1\}^{n-1}\rightarrow \{0,\hdots,2^{n-1}-1\},$ $g\in\{e,o\}$ as \begin{eqnarray*}
   && \alpha_{\Lambda_x}^g(m) =  \sum_{k\in \Lambda_x} m_k2^{k+1}+\sum_{j\not\in \Lambda_x } m_j2^{j+1}+2, \\ &&
   \beta_{\Lambda_x}^e (m)= \sum_{k\in \Lambda_x} \overline{m}_k2^{k+1}+\sum_{j\not\in \Lambda_x} m_j2^{j+1}+2, \\
    && \beta_{\Lambda_x}^o (m)= \sum_{k\in \Lambda_x} \overline{m}_k2^{k+1}+\sum_{j\not\in \Lambda_x} m_j2^{j+1}+1,
\end{eqnarray*} with $m=(m_{n-2},\hdots,m_0)$ and $\overline{m}_k=m_k\oplus 1.$ Then $$\Pi\mathsf{T}^g_{n,x}=\prod_{m=0,\alpha^g_{\Lambda_x(m)}< \beta^g_{\Lambda_x}(m)}^{2^{n-1}-1} P_{(\alpha^g_{\Lambda_x}(m),\beta^g_{\Lambda_x}(m))}, \,\, g\in\{e,o\}.$$ Further  $\Pi\mathsf{T}^g_{n,x}\neq \Pi\mathsf{T}^g_{n,y}$ if $x\neq y$  and $(\alpha^g_{\Lambda_x}(m),\beta^g_{\Lambda_x}(m))\neq (\alpha^g_{\Lambda_y}(m),\beta^g_{\Lambda_y}(m))$ for all $0\leq m\leq 2^{n-1}-1$.

\end{theorem}

For example, if $n=3$ the permutations are defined on the set $\{1,2,\hdots,8\}$. Denote $P_{(i,j)}$ as the permutation matrix corresponding to a two-cycle or transposition $(i,j),$ $i,j\in \{1,2,\hdots,8\}.$ Then the circuits corresponding to $\Pi\mathsf{T}^e_{3,1}, \Pi\mathsf{T}^e_{3,2}$ and $\Pi\mathsf{T}^e_{3,3}$ are given by respectively.

\begin{eqnarray*}
    \Pi\mathsf{T}^e_{3,1}&:=&\Qcircuit @C=1em @R=.7em {&\lstick{1}&\qw&\qw &\qw&\qw&\qw\\
    &\lstick{2}&\qw&\targ &\qw&\qw&\qw\\ &\lstick{3}&\qw&\ctrl{-1} &\qw&\qw&\qw\\}\\
    \Pi\mathsf{T}^e_{3,2}&:=&\Qcircuit @C=1em @R=.7em {&\lstick{1}&\qw&\targ &\qw&\qw&\qw\\
    &\lstick{2}&\qw&\qw &\qw&\qw&\qw\\ &\lstick{3}&\qw&\ctrl{-2} &\qw&\qw&\qw\\}\\
    \Pi\mathsf{T}^e_{3,3}&:=&\Qcircuit @C=1em @R=.7em {&\lstick{1}&\qw&\targ &\qw&\qw&\qw\\
    &\lstick{2}&\qw&\qw &\targ&\qw&\qw\\ &\lstick{3}&\qw&\ctrl{-2} &\ctrl{-1}&\qw&\qw\\}
\end{eqnarray*}
    In particular, we have  \begin{eqnarray*}
        &&\Pi\mathsf{T}^e_{3,1}=P_{(2,4)}P_{(6,8)}, \Pi\mathsf{T}^e_{3,2}=P_{(2,6)}P_{(4,8)},\\&&\Pi\mathsf{T}^e_{3,3}=P_{(2,8)}P_{(4,6)}.
    \end{eqnarray*}
Similarly, the circuits corresponding to $\Pi\mathsf{T}^o_{3,1}, \Pi\mathsf{T}^o_{3,2}$ and $\Pi\mathsf{T}^o_{3,3}$ are given by respectively 
\begin{eqnarray*}
 \Pi\mathsf{T}^o_{3,1}&:=&\Qcircuit @C=1em @R=.7em {&\lstick{1}&\qw&\qw &\qw&\qw&\qw\\
    &\lstick{2}&\ctrl{1}&\targ &\ctrl{1}&\qw&\qw\\ &\lstick{3}&\targ&\ctrl{-1} &\targ&\qw&\qw\\} \\
     \Pi\mathsf{T}^o_{3,2}&:=&\Qcircuit @C=1em @R=.7em {&\lstick{1}&\ctrl{2}&\targ &\ctrl{2}&\qw&\qw\\
    &\lstick{2}&\qw&\qw &\qw&\qw&\qw\\ &\lstick{3}&\targ&\ctrl{-2} &\targ&\qw&\qw\\} \\
     \Pi\mathsf{T}^o_{3,3}&:=&\Qcircuit @C=1em @R=.7em {&\lstick{1}&\ctrl{2}&\targ &\qw&\ctrl{2}&\qw\\
    &\lstick{2}&\qw&\qw &\targ&\qw&\qw\\ &\lstick{3}&\targ&\ctrl{-2} &\ctrl{-1}&\targ&\qw\\}    
\end{eqnarray*}

Similarly, we have \begin{eqnarray*}
    &&\Pi\mathsf{T}^o_{3,1}=P_{(2,3)}P_{(6,7)}, \Pi\mathsf{T}^o_{3,2}=P_{(2,5)}P_{(4,7)},\\&&\Pi\mathsf{T}^o_{3,3}=P_{(2,7)}P_{(4,5)}.
\end{eqnarray*}  

\subsection{Permutation-similarity of Pauli-strings}

In this section, we provide a circuit model to establish a similarity relation between Pauli-strings containing $X$ gate and $I_2$ through permutation matrices $\Pi\mathsf{T}^g_{n,x},$ as defined in Theorem \ref{constructeven}. First we introduce the following notations.

For any integer $b\in\{0,\hdots, 2^n-1\}$ consider the binary representation of $b$ as $(b_{1}, \hdots, b_n),$ $b_j\in\{0,1\}$ such that $b=\sum_{j=1}^{n} 2^{n-j}b_j.$ Then corresponding to any such $b,$ define the Pauli-string matrix $\sigma_b=\otimes_{j=1}^{n} \sigma_{j}=\sigma_1\otimes\hdots\otimes\sigma_n$, where $$\sigma_j=\begin{cases}
    X \,\, \mbox{if} \,\, b_j=1 \\
    I_2 \,\, \mbox{if} \,\, b_j=0.
\end{cases}$$ This notation enables to define an ordering on the set $S_{I_2,X}^{(n)}=\{\sigma_b | 0\leq b\leq 2^n-1\}$ based on the standard ordering of the set $\{0,\hdots,2^n-1\}$ defined by the map $\eta_n: S_{I_2,X}^{(n)} \rightarrow  \{0,1,\hdots,2^n-1\}$ as $\eta_n(\sigma_b)=b.$ Besides, the canonical basis of the $n$-qubit Hilbert space can be written as $\{\ket{b} : 0\leq b\leq 2^n-1\}=\{\ket{b_{1}\hdots b_n} | b_j\in\{0,1\}\}.$

Now we have the following theorem, which establishes permutation similarity of Pauli-strings defined by Pauli matrices  $I_2, X$.

\begin{theorem}\label{SRBBvPauli}
Let $b\in \{1,\hdots,2^{n}-1\}.$ Then  $$\sigma_b=\begin{cases}
    \Pi\mathsf{T}_{n,\frac{b-1}{2}}^e (I_2^{\otimes (n-1)}\otimes X) \Pi\mathsf{T}_{n,\frac{b-1}{2}}^e, \,\, \mbox{if}\,\, b \,\, \mbox{is odd} \\
    \Pi\mathsf{T}_{n,\frac{b}{2}}^o (I_2^{\otimes (n-1)}\otimes X) \Pi\mathsf{T}_{n,\frac{b}{2}}^o, \,\, \mbox{if}\,\, b \,\, \mbox{is even.}
\end{cases}$$

\end{theorem}

\pf See Appendix \ref{appdx:A1}. \hfill{$\square$}

\begin{remark}
    Note that \begin{eqnarray*}
    \hspace{-0.65cm}&&\left|\bigcup_{g\in \{e,o\}}\{\Pi\mathsf{T}_{n,x}^g : 1\leq x\leq 2^{n-1}-1\}\cup I_{2^n}\right|\\=&&\sum_{g\in \{e,o\}}\left|\{\Pi\mathsf{T}_{n,x}^g : 1\leq x\leq 2^{n-1}-1\}\right| +1\\=&&2^n-1
\end{eqnarray*} This is due to the fact that $\{\Pi\mathsf{T}_{n,x}^e : 1\leq x\leq 2^{n-1}-1\}\cap \{\Pi\mathsf{T}_{n,x}^o : 1\leq x\leq 2^{n-1}-1\}=\emptyset$. Further, from \cite{SarmaSarkar2023}, it is to be noticed that all permutations from $\{\Pi\mathsf{T}_{n,x}^e : 1\leq x\leq 2^{n-1}-1\}\cup\{\Pi\mathsf{T}_{n,x}^o : 1\leq x\leq 2^{n-1}-1\}\cup I_{2^n}$ are distinct even in their components i.e. no two components of the transpositions match because each permutation matrix has unique circuit construction. Also $|\mathcal{S}_{I,X}^{(n)} \setminus \{I\}|=2^n-1$. Hence, there is a 1-1 correspondence between $ \{\Pi\mathsf{T}_{n,x}^e : 1\leq x\leq 2^{n-1}-1\}\cup\{\Pi\mathsf{T}_{n,x}^o : 1\leq x\leq 2^{n-1}-1\}\cup I_{2^n}\}$ and $\mathcal{S}_{I,X}^{(n)} \setminus \{I\}$ through the operation ${\Pi\mathsf{T}_{n,{x}}^g}(I_2^{\otimes n-1}\otimes X){\Pi\mathsf{T}_{n,{x}}^g}, g\in \{e,o\}$. \hfill $\square$
\end{remark}  

We also recall that $HXH=Z$ and $Y=S^\dagger XS$ where $$H=\frac{1}{\sqrt{2}}\bmatrix{1&1\\1&-1},\,\, \mbox{and}\,\, S=\bmatrix{e^{\iota\pi/4} &0 \\ 0 & e^{-\iota\pi/4}}=R_Z(\pi/4),$$ $\dagger$ denotes conjugate transpose of a matrix and $R_Z(\theta)=\bmatrix{\exp{(\iota \theta)}&0\\0&\exp{(-\iota \theta)}}$ denotes the rotation gate of a qubit around $Z$-axis by an angle $\theta.$ Then we have the  following theorem.

\begin{theorem}\label{genralPauli}
For any $n$-length Pauli string $\sigma_{1}\otimes \sigma_{2}\hdots\otimes \sigma_n$, $\sigma_j\in\{I,X,Y,Z\}$ the following relation holds  \begin{eqnarray*}
 \hspace{-0.5cm}   (\sigma_{1}\otimes \sigma_{2}\hdots\otimes \sigma_n)&=&(\tau_{1}\otimes \tau_{2}\hdots\otimes \tau_n)^\dagger(\mu_{1}\otimes \mu_{2}\hdots\otimes \mu_n)\\&&(\tau_{1}\otimes \tau_{2}\hdots\otimes \tau_n)
\end{eqnarray*}where $$\begin{cases}
    \tau_j=H,\mu_j=X,\mbox{ if } \sigma_j=Z\\
     \tau_j=S,\mu_j=X,\mbox{ if } \sigma_j=Y\\
     \tau_j=I,\mu_j=X,\mbox{ if } \sigma_j=X\\
     \tau_j=I,\mu_j=I,\mbox{ if } \sigma_j=I,
\end{cases}$$  $j\in \{0,\hdots,n-1\}.$  
\end{theorem}
\pf Follows from elementary properties of Kronecker product, $HXH=Z,$ and $Y=S^\dagger XS$.\hfill $\square$

\subsection{Design of scalable quantum circuits for exponential of Pauli-strings}

In this section, we shall provide a scalable quantum circuit model representation of Pauli-string exponentials using only $1$-qubit rotation gates and $\cnot$ gates.

It is obvious to see that circuit for $\exp{(i\theta I_2^{\otimes (n-1)}\otimes X )}$ is \begin{eqnarray} \label{expoIX}
{\Qcircuit @C=1em @R=.7em{
 &\lstick{1}&\qw&\qw&\qw&\qw\\
 &\lstick{2}&\qw&\qw&\qw&\qw\\
 &\lstick{\vdots}&\qw&\qw&\qw&\qw\\
 &\lstick{n}&\qw&\gate{R_X(\theta)}&\qw&\qw\\}} 
\end{eqnarray} where $R_X(\theta)=\bmatrix{\cos{\theta} &\iota \sin{\theta}\\\iota\sin{\theta}&\cos{\theta}}$ denotes the qubit's rotation gate around the $X$-axis by an angle of $\theta$.

Now,  we have  the following Theorem. 
\begin{theorem}\label{pauliexp}
    Given any $n$-length Pauli string $\sigma\in \mathcal{S}^{(n)}_{I,X}\setminus \{I_{2^n}\}$. Then the circuit for $\exp{(i\theta \sigma)}$ is given as
        \begin{eqnarray*}\label{circpauliexp}
  \Qcircuit @C=1em @R=.7em {
    &\lstick{1}&\multigate{3}{P}& \qw &\multigate{3}{P}&\qw&\qw\\
    &\lstick{\vdots}&\ghost{P}& \qw&\ghost{P}&\qw &\vdots\\
    &\lstick{n-1}&\ghost{P}&\qw&\ghost{P}&\qw&\qw\\
    &\lstick{n}&\ghost{P}&\gate{R_X(\theta)}&\ghost{P}&\qw&\qw\\}\end{eqnarray*}where $P\in\{\Pi\mathsf{T}_{n,x}^g : g\in\{e,o\}, 0\leq x\leq 2^{n-1}-1 \}$ is a permutation matrix  mentioned Theorem \ref{SRBBvPauli} such that $P\sigma P=I_2^{\otimes (n-1)}\otimes X$.
\end{theorem}
\pf From Theorem\ref{SRBBvPauli}, we see that any Pauli string $\sigma\in \mathcal{S}^{(n)}_{I,X}\setminus \{I\}$ can be written as $P\sigma P$ for some symmetric permutation matrix $P$ where $\sigma^{(n)}=I_2^{\otimes (n-1)}\otimes X$. Clearly, from standard linear algebra, $\exp{(\iota \theta \sigma)}=P\exp{(\iota \theta \sigma)} P$. The rest of the proof follows immediately.  \hfill $\square$

Finally, we show how to express exponential of any scaled $n$-length Pauli-string in a quantum circuit.

\begin{corollary}\label{pauliexpgen}
    Given any $n$-length Pauli string such that $\sigma=(\sigma_{1}\otimes \sigma_2\hdots\otimes \sigma_n)\neq I$. Then $\sigma$ can be written in the form  $(\sigma_{1}\otimes \sigma_{2}\hdots\otimes \sigma_n)=(\tau_{1}\otimes \tau_{2}\hdots\otimes \tau_n)^\dagger(\mu_{1}\otimes \mu_{2}\hdots\otimes \mu_n)(\tau_{1}\otimes \tau_{2}\hdots\otimes \tau_n)$ where $\tau_j,\mu_j, j\in\{1,\hdots,n\}$ are Pauli matrices satisfying the conditions in Theorem \ref{genralPauli}. Then the circuit for $\exp{(\pm \iota \theta \sigma)}$ is given as
 \small\begin{eqnarray}\label{circpauliexpcor}
 {\Qcircuit @C=1em @R=.7em {
    &\lstick{1}&\gate{\tau_{1}}&\multigate{3}{P}& \qw &\multigate{3}{P}&\qw&\gate{\tau_{1}^\dagger}&\qw\\
    &\lstick{\vdots}&\qw &\ghost{P}&\qw &\ghost{P}&\qw&\qw &\vdots\\
    &\lstick{n-1}&\gate{\tau_{n-1}}&\ghost{P}&\qw&\ghost{P}&\qw&\gate{\tau_{n-1}^\dagger}&\qw\\
    &\lstick{n}&\gate{\tau_{n}}&\ghost{P}&\gate{R_X(\pm\theta)} &\ghost{P}&\qw&\gate{\tau_n^\dagger}&\qw\\}}\end{eqnarray}\normalsize where $P$ is a permutation matrix derived from Theorem\ref{SRBBvPauli} such that $P(\mu_{1}\otimes \mu_{2}\hdots\otimes \mu_n)P=I_2^{\otimes (n-1)}\otimes X$. 
\end{corollary}
\pf The proof follows from Theorem \ref{genralPauli} and Theorem \ref{pauliexp}. Indeed, note that $\sigma=\tau \mu \tau$ implies $\exp(\pm \iota \theta \sigma)=\tau\exp(\pm\iota \mu)\tau$ since $\tau=\otimes_{j=1}^n\tau_j$ is unitary and $\mu=\otimes_{j=1}^n\mu_j.$ \hfill $\square$ 

Now we provide Algorithm \ref{Pauliblock} for the construction of exponential of $n$-length Pauli-strings as discussed in the theorems above.

We formulate Algorithm \ref{Pauliblock} that helps us to convert any Pauli string in $S_n\setminus I_{2^n}$ to $I^{\otimes (n-1)}\otimes X.$ Using Algorithm\ref{Pauliblock}, it follows that  deriving the matrix representation of elements of $\mathcal{S}_{I,X}^{(n)}$ can be performed in $O(2^{n-1}).$
From the circuit of Pauli string exponentials, we prove the following theorem that showcases total number of $\cnot$ and one-qubit rotation gates required for the circuit model.

\begin{theorem}\label{gatecount}
  Let $\sigma$ be a $n$-length Pauli-string and let there be $n_1,$ $n_2$, $n_3$ and $n_4$ number of $X$, $Y,$ $Z$ and $I_2$ gates respectively that comprise $\sigma$ and $n=n_1+n_2+n_3+n_4$. Then, the quantum circuit of $\exp{(\iota\theta\sigma)}$ requires at most $1$ $R_X$ gate, $2n_2$ $R_Y$ gates, $4n_2+2n_3$ $R_Z$ gates, $2n-2$ gates of the form $\cnot_{n,l}$ and atmost $2$ gates of the form $\cnot_{l,n}$ where $l\in \{1,\hdots,n-1\}$. If $\sigma\in \mathcal{S}^{(n)}_{I,Z}$, then at most $2n-2$ gates of the form $\cnot_{l,n}$ and $1$ $R_Z$ gates are required to construct $\exp{(\iota\theta\sigma)}$.    
\end{theorem}
\pf From the construction of exponential of Pauli strings in Algorithm \ref{Pauliblock} and Corollary \ref{pauliexpgen},  we see that all permutation matrices $\Pi\mathsf{T}_{n,x}^g,1\leq x\leq 2^{n-1}-1,g\in \{e,o\}$ require at most $n+1$ $\cnot$ gates. Such a bound is obtained for $\Pi\mathsf{T}_{n,2^{n-1}-1}^o$, where $n-1$ $\cnot_{l,n}$ gates and $2$ $\cnot_{1,n}$ are required (See Theorem \ref{constructeven} or \cite{SarmaSarkar2023}). Since such permutation matrices are multiplied on both sides of $I_2^{\otimes (n-1)}\otimes X$, then at most $2n-2$ $\cnot_{n,l}$ and $4$ $\cnot_{l,n}$ gates are required. Now, it is easy to see that the circuits  \\\centerline{\Qcircuit @C=1em @R=1.3em {
    &\lstick{1}&\qw&\qw&\qw&\qw&\qw\\
    &\lstick{2}&\qw&\qw&\qw&\qw&\qw\\
    &\lstick{\vdots}&\qw&\qw&\qw&\qw&\qw\\
    &\lstick{l}&\qw&\ctrl{3}&\qw&\ctrl{3}&\qw\\
    &\lstick{\vdots}&\qw&\qw&\qw&\qw&\qw\\
    &\lstick{n-1}&\qw&\qw&\qw&\qw&\qw\\
    &\lstick{n}&\qw&\targ&\gate{R_X(\theta)}&\targ&\qw\\}} and\\ \centerline{\Qcircuit @C=1em @R=1.3em {
    &\lstick{1}&\qw&\qw&\qw&\qw&\qw\\
    &\lstick{2}&\qw&\qw&\qw&\qw&\qw\\
    &\lstick{\vdots}&\qw&\qw&\qw&\qw&\qw\\
    &\lstick{l}&\qw&\qw&\qw&\qw&\qw\\
    &\lstick{\vdots}&\qw&\qw&\qw&\qw&\qw\\
    &\lstick{n-1}&\qw&\qw&\qw&\qw&\qw\\
    &\lstick{n}&\qw&\qw&\gate{R_X(\theta)}&\qw&\qw\\}} are equivalent. 
    
    Hence, at most $2$ $\cnot_{l,n}$ are sufficient where $l\in\{1,\hdots,n-1\}$. Further from Corollary \ref{pauliexpgen}, and Algorithm \ref{Pauliblock} each $\tau_j,0\leq j\leq n-1$ require $2$ $R_Z$ gates and $1$ $R_Y$ gates when $\tau_j=H$. When $\tau_j=S=R_Z(\pi/4)$, only one $R_Z$ is needed. Such single-qubit gates are multiplied both sides i.e. added to both ends of the quantum circuit. Hence, the result follows.   \hfill $\square$

\begin{algorithm}[htb!]
\caption{Constructing quantum circuits of any $n$-length Pauli string exponentials}\label{Pauliblock} 
 \begin{enumerate}
 \item[]\textbf{Provided:}
    \item $n$-length Pauli strings that is not $I^{\otimes n}$ 
    \item $\{\Pi\mathsf{T}_{n,x}^e|1\leq x\leq 2^{n-1}-1\},\{\Pi\mathsf{T}_{n,x}^o|1\leq x\leq 2^{n-1}-1\}$ as defined in equation \ref{permus1} and Theorem\ref{constructeven}
    \item Hadamard gate $H$ and $S=R_Z(\pi/4)$
\end{enumerate} 
\textbf{Input:} A $n$-length Pauli string $\sigma=\sigma_1\otimes \hdots\sigma_n\neq I^{\otimes n}$ where $\sigma_j\in \{I,X,Y,Z\},j\in \{1,\hdots,n\}$.
\begin{enumerate}
    \item[]\textbf{Output:} Quantum circuit for $\exp{(\pm i\theta\sigma)}$
\end{enumerate}
\begin{algorithmic}
\State $J\rightarrow\emptyset$
\State $J_1\rightarrow\emptyset$
\State $J_2\rightarrow \emptyset$
\For{$k=1:n:k++$}
\State $\mu_k\rightarrow I$
\State $\tau_k\rightarrow I$
\State End For
\EndFor
\For{$k=1:n:k++$}
\If{$\sigma_k==X$}
\State $J\rightarrow J\cup\{k\}$
\ElsIf{$\sigma_k==Y$}
\State $J_1\rightarrow J_1\cup\{k\}$
\ElsIf{$\sigma_k==Z$}
\State $J_2\rightarrow J_2\cup\{k\}$
\State End If
\EndIf
\State End For
\EndFor
\For{$k=1:n:k++$}
\If{$k\in J$ and $k\in J_1$}
\State $\mu_k\rightarrow X$
\State End If
\EndIf
\State End For
\EndFor
\State $\mu\rightarrow\mu_1\otimes \hdots\otimes \mu_n$
\State $b\rightarrow \sum_{k\in J\cup J_1}^{n} 2^{n-k}$
\If{$b$ is odd}
\State $x\rightarrow \frac{b-1}{2}$
\State $P\rightarrow \Pi\mathsf{T}_{n,x}^e$
\Else
\State $x\rightarrow \frac{b}{2}$
\State $P\rightarrow \Pi\mathsf{T}_{n,x}^o$
\State End If
\EndIf
\For{$k=1:n:k++$}
\If{$k\in J_1$}
\State $\tau_k\rightarrow S$
\ElsIf{$\sigma_k==Z$}
\State $\tau_k\rightarrow H$
\State End If
\EndIf
\State End For
\State \textbf{Output: Circuit \ref{circpauliexpcor}}
\EndFor
\State End
\end{algorithmic}
\end{algorithm} 


It is to be noted that there are several approaches proposed in the literature to design and implement exponential of scaled Pauli-strings through quantum circuits \cite{Raeisi2012, Reiher2017,vanden2020, Yordanov2020, Clinton2021, Kawase2022, Mukhopadhyay2023, Mansky2023}. Generally, such methods are coupled with Suzuki-Trotter decomposition (or Trotterization in short) in order to simulate diverse classes of Hamiltonians through quantum circuits. For a detailed description of Trotterization, refer to Appendix \ref{qc_hamil}. The circuit methods in literature along with the method proposed in this paper can be compared dealing with two properties of the circuits: scalability and the assumption of fully connected quantum hardware. 

It is evident from our proposed quantum circuit design that it is scalable and it does not need of complete or dense connectivity of a quantum hardware for implementing the two-qubit controlled gates in the circuit owing to the permutation similarity of certain Pauli-strings. These overcome some of the challenges in circuit implementation of $\exp(\iota\theta\sigma)$ for any Pauli-string $\sigma$ and $\theta\in \R$ as confronted in the exisiting literature. It should also be emphasized that the existing quantum hardware architecture  such as  IBM Q5 Yorktown/IBM QX2 quantum architecture for $5$ qubits \cite{Ibm2017} support the implementation of the proposed circuit model for exponential of non-diagonal Pauli-strings due to its requirement of implementation of controlled gates such that the control qubit is the $n$-th qubit. For exponential of diagonal Pauli-strings i.e. Pauli-strings formed by $I_2$ and $Z$ require the implementation of control gates with target qubit as the $n$-th qubit (see Theorem \ref{gatecount}).


\begin{figure*}[ht]
    \subfigure{\includegraphics[width=0.45\textwidth]{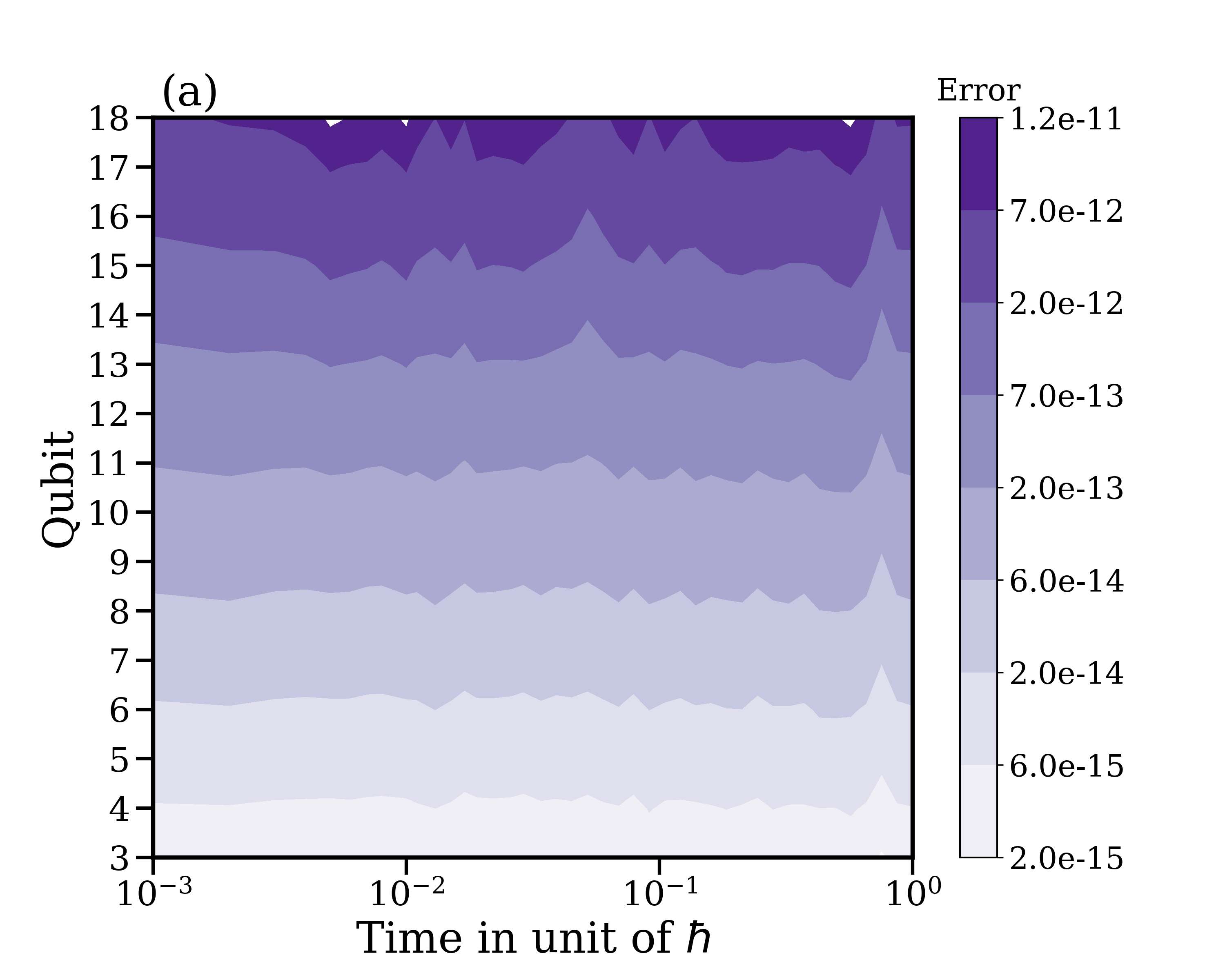}}
     \subfigure{\includegraphics[width=0.45\textwidth]{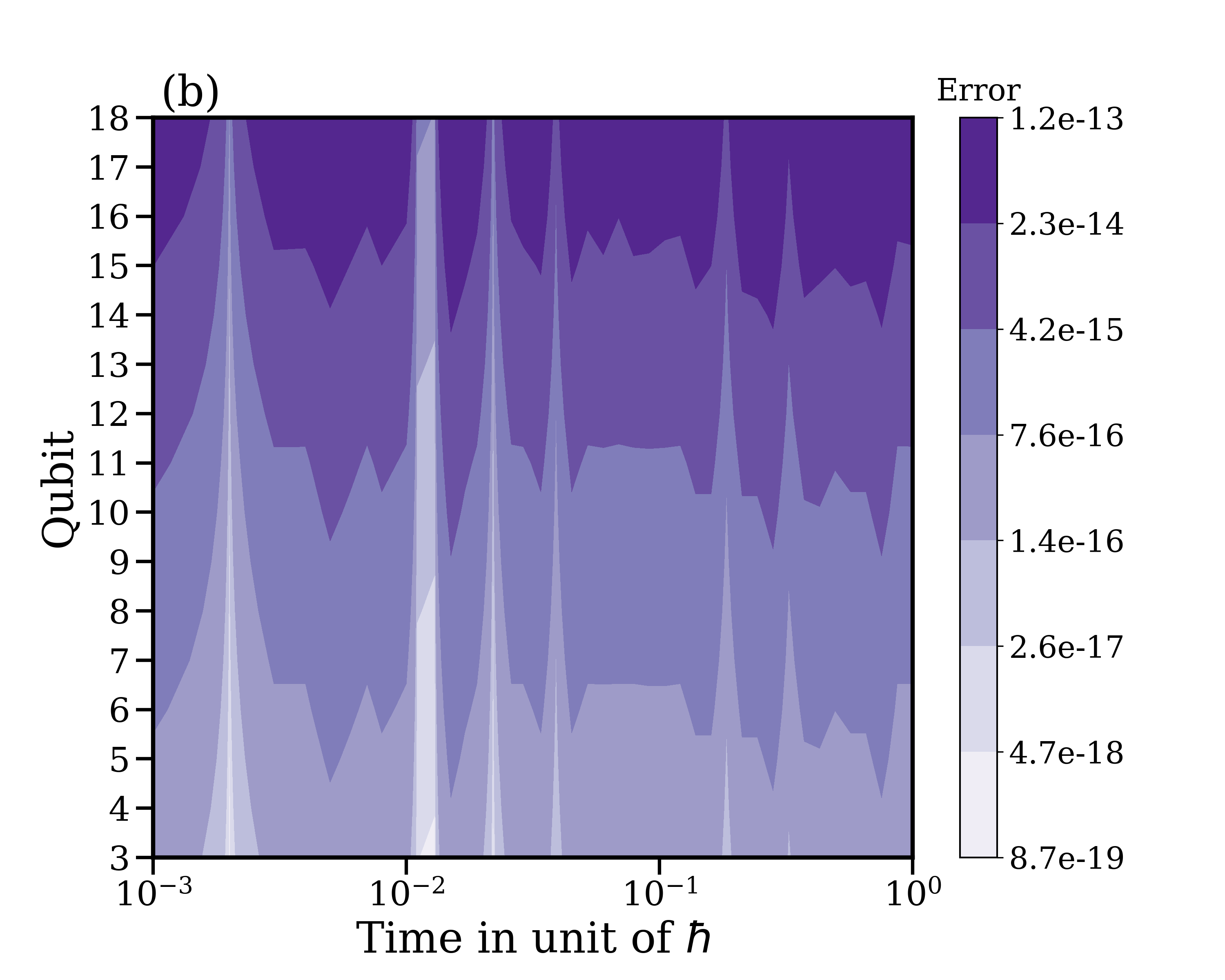}}
    \caption{Surface plot of circuit simulation error for fixed depth Hamiltonian simulation of (a) $1$-sparse Quantum Ising Hamiltonian and (b) $2$-sparse Hamiltonian corresponding to a $2 \times 2$ block-diagonal unitary respectively. In the figure the time is measured in units of $\hbar$ and the simulation is done till 18 qubit. The time evolution spans total time $T=1.0$, with each time step set at $t=0.001$. The ``Error" in the colourbar is quantified as the Frobenius norm distance between the actual unitary evolution corresponding to the Hamiltonian operator and the unitary matrix obtained through the circuit model. We observe that the error value rises with an increase in Qubit number $n$ for both cases.
    }\label{fig:fixeddepth}
\end{figure*}

\section{Quantum circuit simulation of Hamiltonian operators}\label{hsim_qcktall}

In this section, we report quantum circuit approximation errors corresponding to Frobenius norm for the unitary evolution $U(t)$ defined by several classes of 1D Hamiltonian operators utilizing the proposed quantum circuit representation of $\exp(\iota\theta\sigma)$ by employing Suzuki-Trotter approximation formula. We consider $2$-sparse block-diagonal Hamiltonian ($\mathcal{H}_{Blkdg}$), Ising Hamiltonian ($\mathcal{H}_{IM}$), and both time-independent and time-dependent Random Field Heisenberg Hamiltonian ($\mathcal{H}_{Heis}$) and Transverse Magnetic Random Quantum Ising Hamiltonian ($\mathcal{H}_{TFQIM}$) using first order Suzuki-Trotter formula. We further examine the performance of the proposed quantum circuit simulation using higher-order Trotterization approximation formula considering a token Hamiltonian:  time-independent transverse Magnetic Field Quantum Ising Hamiltonian. We observe that circuit approximation error closely matches with the numerical Trotter approximation error for all the performed simulations for up to $18$-qubit systems. 


The mathematical descriptions of the Hamiltonian operators, written as linear combination of Pauli strings are given below. The matrices corresponding to these Hamiltonian operators are sparse and the sparsity pattern plays an important role during the simulations. We call a matrix $k$-sparse if each row and column has at most $k$ non-zero entries. Obviously, Pauli-strings are $1$-sparse Hermitian unitary matrices. We also identify a salient feature of the proposed quantum circuit model in simulating certain Hamiltonian operators that the proposed circuits have constant depth, see Section \ref{fixed_depth_sim}. Finally, we incorporate the noise models as described in Section \ref{Noise_Model_details} in order to investigate the efficiency of the proposed circuit model for Hamiltonian simulation in a NISQ like environment. 

All simulations are carried out on Qiskit \cite{Qiskit} on a system with 16GB RAM, 12th Gen Intel(R) Core(TM) i5-1235U; 1.30 GHz till $10$-qubit systems. Beyond $10$-qubit systems, we have performed the simulations using the supercomputer PARAM Shakti of IIT Kharagpur, established under National Supercomputing Mission (NSM), Government of India and supported by Centre for Development of Advanced Computing (CDAC), Pune \cite{Paramshakti}.


\begin{figure*}[ht]
\subfigure{\includegraphics[width=0.45\textwidth]{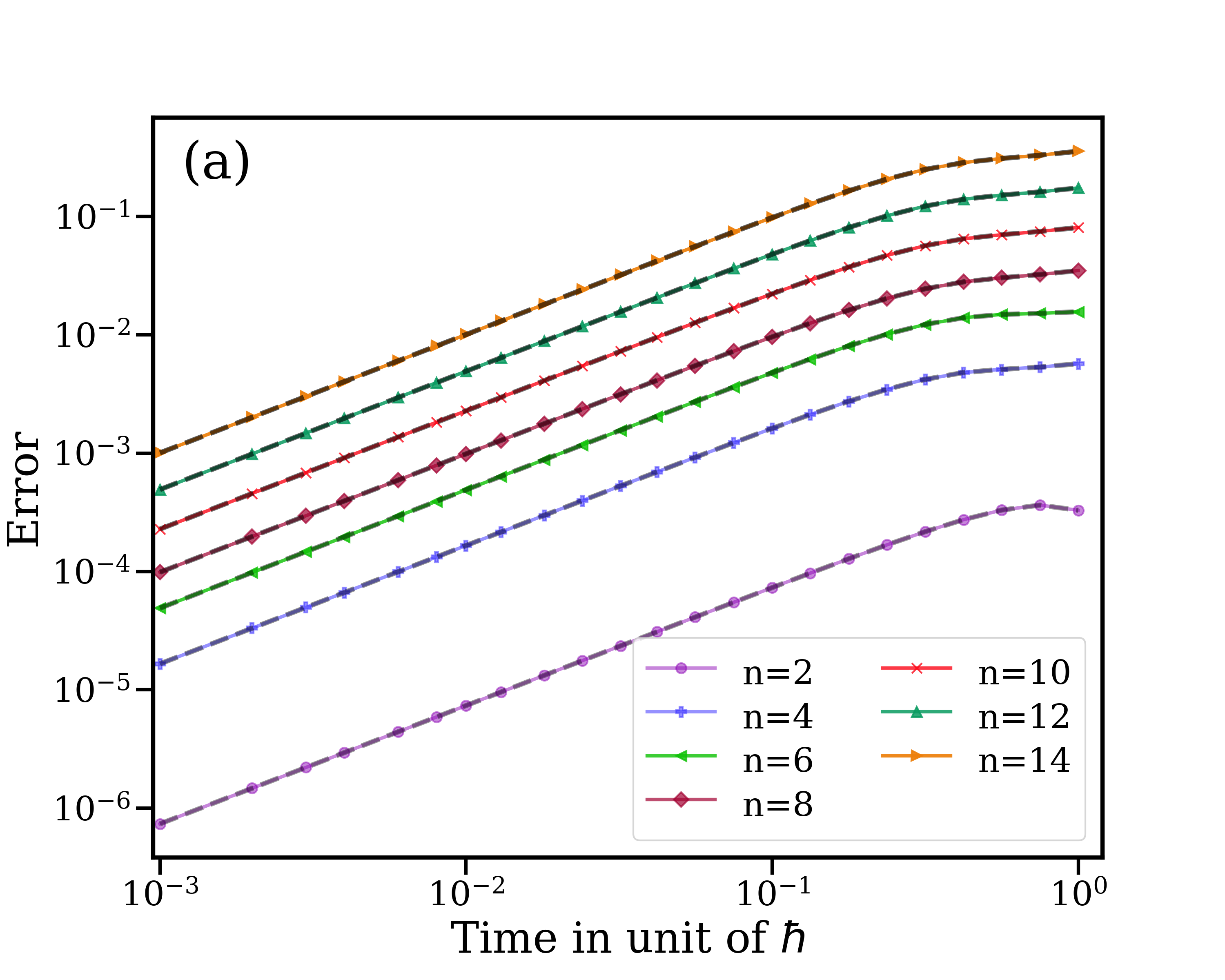}}
   \subfigure{\includegraphics[width=0.45\textwidth]{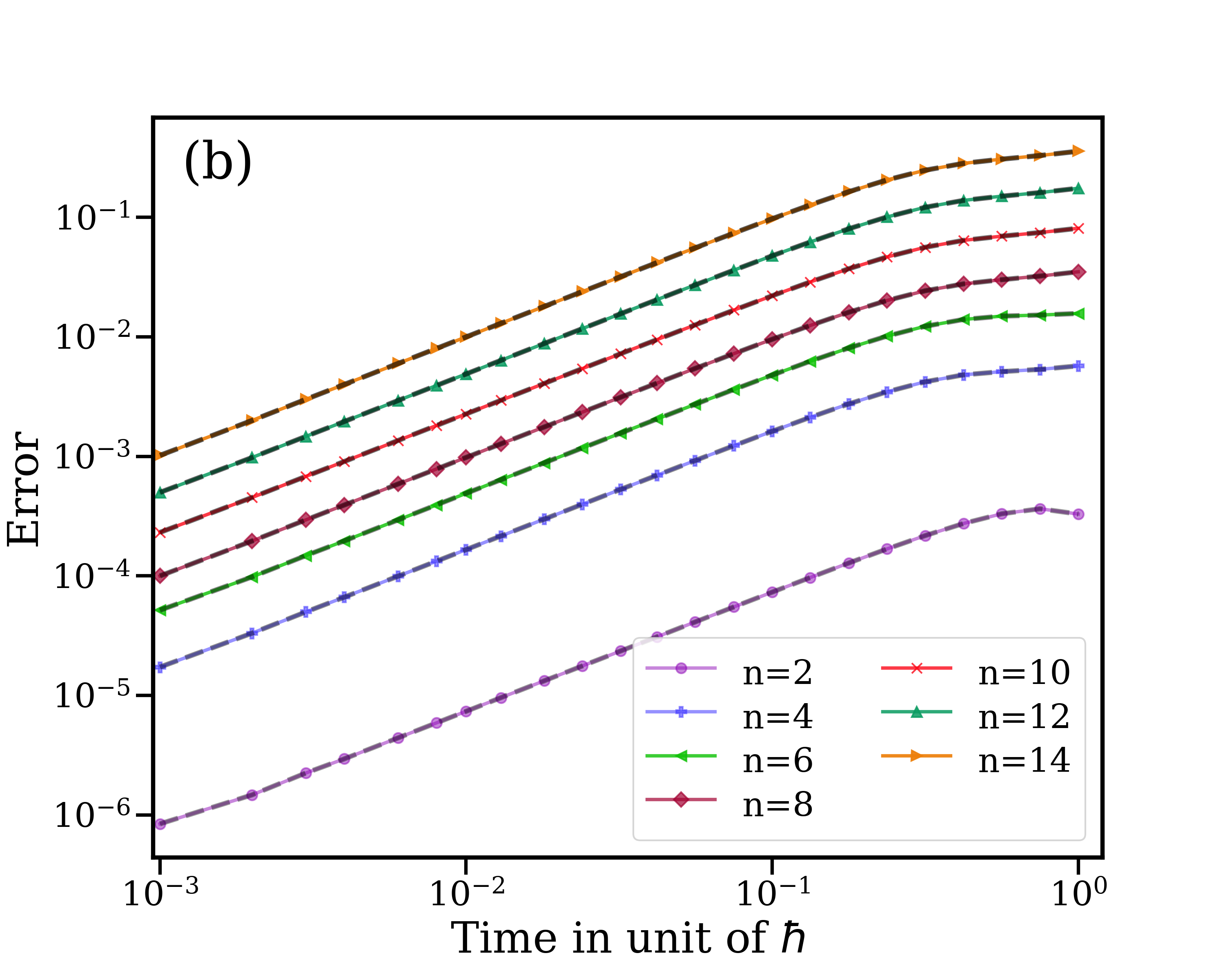}}
   \caption{Line plot of circuit simulation error for Random field Heisenberg Hamiltonian $\mathcal{H}_\textsc{HEIS}$ for the (a) time-independent and (b) time-dependent case. Time is measured in units of $\hbar$. The time evolution spans $T=1.0$, with each time step set at $\frac{T}{r}=0.001$. The plot illustrates the numerical Trotter error using solid lines with distinct markers for different qubit counts as follows 2 qubits (violet, $\circ$), 4 qubits (blue, $+$), 6 qubits (green, $\lhd$), 8 qubits (purple, $\diamond$), 10 qubits (red, $\times$), 12 qubits (teal-green, $\bigtriangleup$), and 14 qubits (orange, $\rhd$). The black dotted line depicts the error due to the proposed circuit approach, which closely mirrors the numerical Trotter decomposition error.  
    }\label{Heisenberg_Hamiltonian}
\end{figure*}

\subsection{Circuit simulation with constant circuit depth for $\mathcal{H}_\textsc{IM}$ and $\mathcal{H}_\textsc{Blkdg}$ }\label{fixed_depth_sim}

In this section, we consider quantum circuit approximation for unitary evolution $U(t)$ for $2$-sparse block diagonal and quantum Ising model Hamiltonian operators. We illustrate that the proposed quantum circuits for these Hamiltonians are of constant depth, where the depth of a quantum circuit is  measured as the number of ``layers" of quantum gates that must operate concurrently in the circuit \cite{ibmCircuitQuantum}.


The mathematical expression of quantum Ising model on 1D spin chain is given by \cite{pires2021theoretical} $$\mathcal{H}_\textsc{IM} = \sum_{l=1}^{n-1} J_{l,l+1} Z_{l}Z_{l+1},$$  where $J_{l,l+1}>0$ denotes the interaction between adjacent sites $l$, $l+1$ and $Z_l$ denotes the action of $Z$ on the $l$-th site. Here we consider $J_{l,l+1}=J=1$ for all $l$. It is evident that all local Hamiltonian operators along with their matrix exponential commute with each other on account of being diagonal matrices. Hence, \begin{eqnarray}\label{fixed1}
    \exp{(-\iota \mathcal{H}_\textsc{IM}T)}&=&\left(\exp{(-\iota \mathcal{H}_\textsc{IM})\frac{T}{r}}\right)^r\\\nonumber&=&\left(\prod_{l=1}^{n-1}\exp{\left(-\iota J_{l,l+1}Z_{l}Z_{l+1}\frac{T}{r}\right)}\right)^r\\\nonumber&=&\left(\prod_{l=1}^{n-1}\exp{(-\iota J_{l,l+1}Z_{l}Z_{l+1}T)}\right).
\end{eqnarray} Thus, in the quantum circuits of the Pauli-strings corresponding to diagonal matrices obtained via Algorithm \ref{Pauliblock}, one must change the angle of the rotation gate $R_X\left(\theta\right)$ from $-J_{l,l+1}\frac{T}{r}$ in the first step to $(-kJ_{l,l+1}\frac{T}{r})$ in the $k$-th time step to finally $\left(-rJ_{l,l+1}\frac{T}{r}\right)=(J_{l,l+1}{T})$. Due to the change of this angle in every time-step coupled with the equation (\ref{fixed1}), the number of layers while simulating Hamiltonian operator in $\mathcal{H}_{IM}$ remains constant, which provides us with a fixed depth quantum circuit.


The general form of a  $2$-sparse block-diagonal matrices for the Hamiltonian operators are 
\begin{eqnarray}
    \mathcal{H}_{\textsc{blkdg}}&=&\sum_{\sigma_\alpha\in\{\ I_2,Z\}, \tau_\alpha \in \left \{ I_2,X, Y,Z \right \}}c_{\alpha} \sigma_\alpha^{\bigotimes n-1} \otimes \tau_\alpha,  
\end{eqnarray} where $\alpha\in \Lambda$, some index set and $c_\alpha\in \mathbb{R}$. These Hamiltonians are inspired by the structure of the Jordan-Wigner transformation of the Majorana fermion operators which are mapped as $Z^{\otimes n-1}\otimes X$ and $Z^{\otimes n-1}\otimes Y$, see \cite{kay2016quantum,backens2019jordan}. Here in particular, the Hamiltonians corresponding to $\sigma_\alpha=I_2,\tau_\alpha\in \{I_2,X\}$ and $c_{\alpha}=1$ is taken as a testing block-diagonal Hamiltonian. 
        
For $2$-sparse Hamiltonian operators, we see that $I_2^{\bigotimes n} $ commutes with $I_2^{\bigotimes n-1} \otimes X$. Hence, \small\begin{eqnarray*}
 \exp{(-\iota H_{\textsc{Blkdg}}T)}&=& \left(\exp{\left(-\iota H_{\textsc{Blkdg}}\frac{T}{r}\right)}\right)^r \\
   & =& (\exp{\left(-\iota (I_2^{\bigotimes n})\frac{T}{r}\right)}\\&&\exp{\left(-\iota \left(I_2^{\bigotimes n-1} \otimes X\right)\frac{T}{r}\right)})^r.
\end{eqnarray*}\normalsize
Thus, we obtain \small\begin{eqnarray}\label{fixed2}
     \exp{(-\iota H_{\textsc{Blkdg}}T)}&=& \underbrace{(\exp{(-\iota (I_2^{\bigotimes n})T)}}_{\mbox{diagonal part}}\\&& \underbrace{\left(\exp{\left(-\iota(I_2^{\bigotimes n-1} \otimes X)\frac{T}{r}\right)}\right)^r}_{\mbox{non-diagonal part}}.
\end{eqnarray}\normalsize 

In such cases, as well, we see that the circuit shows fixed depth as one can simply change the angle of $R_X$ obtained from Algorithm \ref{Pauliblock} from $-T/r$ to $-T$ in $r$-steps and using equation (\ref{fixed2}). 

For simulating generic block diagonal matrices, we refer \cite{SarmaSarkar2023, sarkar2024quantum} where the authors have used generalized Z-Y-Z decomposition along with providing a corresponding fixed depth circuit. 
In this way, $\mathcal{H}_{\textsc{Blksdg}}$ produces constant depth circuits for Hamiltonian simulation. Additionally, due to this phenomena, Trotter approximation error is not introduced in our circuit while writing the time-evolution unitary. Figure \ref{fig:fixeddepth} (a) provides the quantum Ising Hamiltonian simulation and Figure \ref{fig:fixeddepth} (b) the $2$-sparse  Hamiltonian simulation following the circuit approach (Section \ref{ckt_model_Pauli_exp}) with time measured in units of $\hbar$. Following equation (\ref{trotter}) in Appendix \ref{qc_hamil}, the time evolution spans from $0$ to $T=1.0$, with each time step set at $\frac{T}{r}=0.001$. As both of the Hamitonian operators contain commutative relationships amongst local Hamiltonian, the Trotter decomposition following equation (\ref{trotter}) is exact making the Trotter error zero. The errors in Figure \ref{fig:fixeddepth}  represent the surface plot of Frobenius norm distance between the actual unitary evolution corresponding to the Hamiltonian operator and the unitary matrix obtained through the circuit model. 

\begin{figure*}[ht]
       \subfigure{\includegraphics[width=0.45\textwidth]{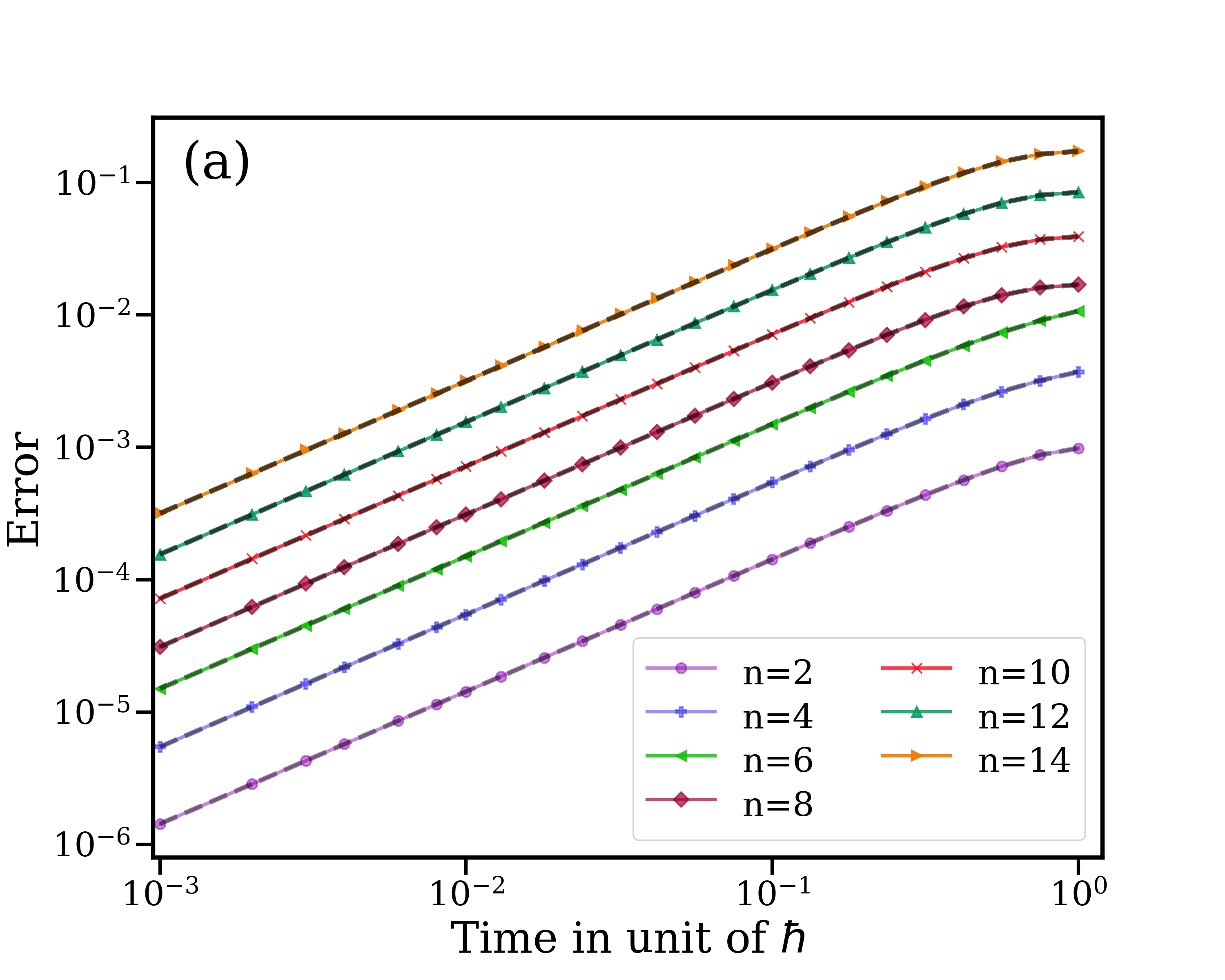}}
     \subfigure{\includegraphics[width=0.45\textwidth]{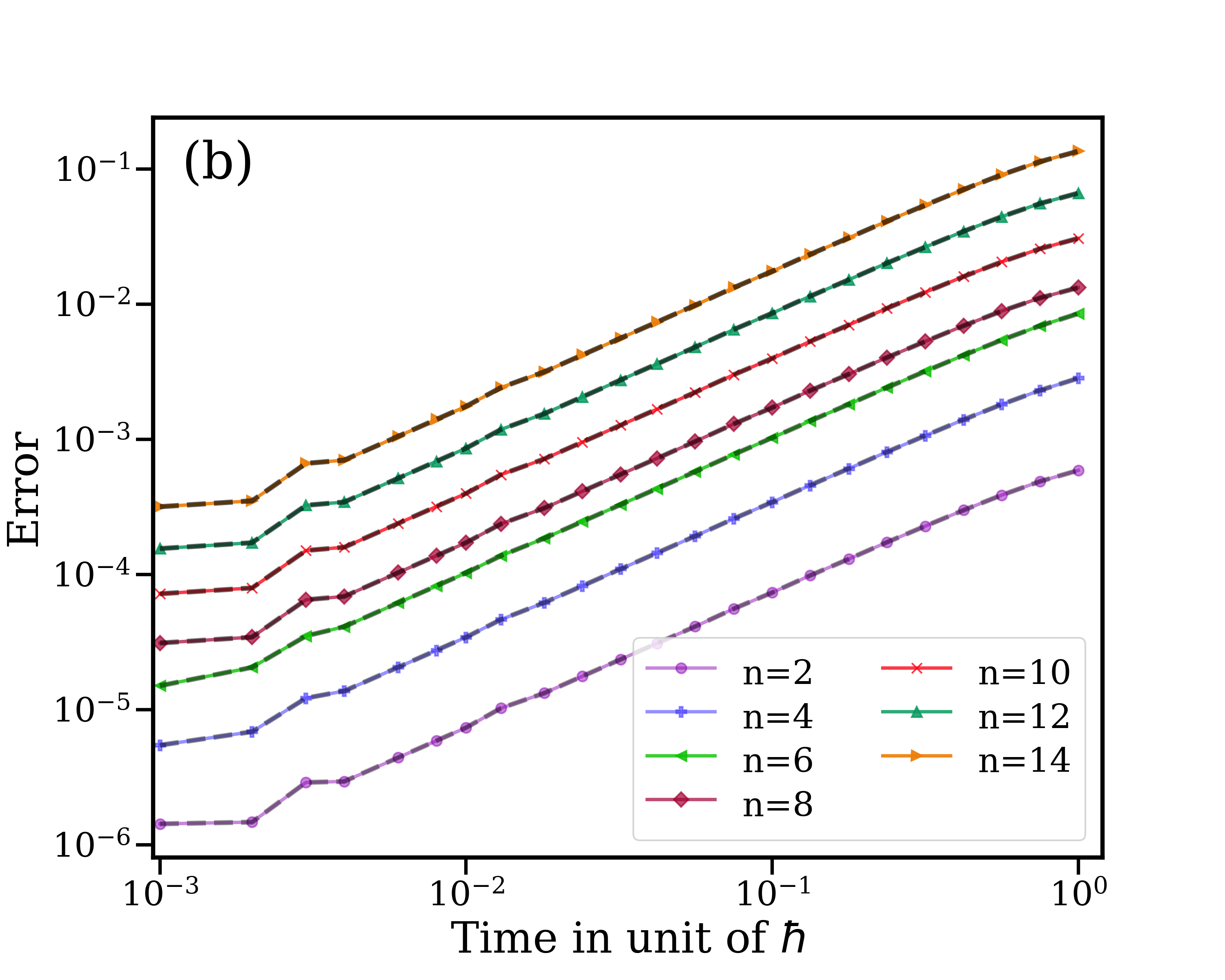}}
  \caption{Line plot for Hamiltonian simulation of the Transverse Magnetic Field Quantum Ising Model (TFQIM) with random field using the Trotter product formula, both in the (a) time-independent and (b) time-dependent cases. Time is measured in units of $\hbar$. The time evolution spans $T=1.0$, with each time step set at $\frac{T}{r}=0.001$. The plot presents the numerical Trotter error with solid lines and distinct markers indicating different qubit counts as follows 2 qubits (violet, $\circ$), 4 qubits (blue, $+$), 6 qubits (green, $\lhd$), 8 qubits (purple, $\diamond$), 10 qubits (red, $\times$), 12 qubits (teal-green, $\bigtriangleup$), and 14 qubits (orange, $\rhd$). The black dotted line represents the error from the proposed circuit method, which closely aligns with the numerical Trotter decomposition error. 
  }  \label{TFQIM_Hamiltonian}
\end{figure*}

\subsection{Circuit simulation of time-independent and time-dependent $\mathcal{H}_\textsc{HEIS}$ and $\mathcal{H}_\textsc{TFQIM}$ }\label{ckt_Heis_TFQIM}

In this section, we consider the quantum circuit approximation for unitary evolution of the operators, Heisenberg Hamiltonian ($\mathcal{H}_\textsc{Heis}$) and transverse magnetic field quantum Ising Hamiltonian ($\mathcal{H}_\textsc{TFQIM}$) for both time-independent and time-dependent cases. Since the local Hamiltonian operators are $1$-sparse matrices, their individual Frobenius norms attain a low magnitude of error ranging from $O(10^{-16})$ to $O(10^{-12})$. Thus, during the time evolution while Trotter decomposition is implemented, the total Hamiltonian simulation error using the circuit approach (black dotted lines) closely mirrors the numerical Trotter decomposition error (solid lines with distinct markers for different qubit counts), represented in Figures \ref{Heisenberg_Hamiltonian}, \ref{TFQIM_Hamiltonian}. 

The Heisenberg Hamiltonian for both time-independent and dependent cases are commonly used in the study of magnetic systems in which magnetic spins are treated quantum mechanically \cite{fazekas1999lecture,skomski2008simple,pires2021theoretical}. Only spin-spin interactions are included in the said Hamiltonian with a random field, which is expressed as
\begin{eqnarray}\label{heis}
    \mathcal{H}_{\textsc{Heis}} &=& \sum_{l=1}^{n-1} J_{l,l+1}\left ( X_{l}X_{l+1} + Y_{l}Y_{l+1} + Z_{l}Z_{l+1} \right )\\\nonumber && + \sum_{l=1}^{n}h_{l}Z_{l},
\end{eqnarray}
where $J_{l,l+1} > 0$ denotes the coupling parameter which specifies the exchange interaction between nearest-neighbour spin. We set  $J_{l,l+1}=J$, a constant. Furthermore, we have taken $J=1$, which depicts the antiferromagnetic case. Besides, $h_{l}$ denotes the uniformly random coefficients which are chosen from a standard normal distribution on $[-1, \, 1]$.

The Hamiltonian can be further modified by invoking certain constraints such as interaction with the magnetic field that depends on time. In such cases the Hamiltonian becomes,
\begin{eqnarray}\label{Heistimedep}
    \mathcal{H}_{\textsc{Heis}}(t) &=& \mathcal{H}_\textsc{Heis} - g(t)\sum_{l=1}^{N}X_{l},
\end{eqnarray}
where $g(t)=\sin{\frac{{\omega(t)}\pi}{2}}$ such that $\omega(t)$ takes the value $1$ or $-1$ depending on even or odd time-steps respectively during unitary evolution.

Along with sparse Hamiltonians, we also investigate the transverse magnetic field quantum Ising model (TFQIM) with nearest-neighbor interactions defined as \cite{pires2021theoretical}
\begin{eqnarray}\label{tfqim}
   \mathcal{H}_\textsc{TFQIM} &=& J\sum_{l=1}^{n-1}Z_{l}Z_{l+1} + g\sum_{l=1}^{n}X_{l} \\\nonumber&& + \sum_{l=1}^{n}h_{l}Z_{l}.
\end{eqnarray}
Here the sum runs over site $l$ and $l$ and $l+1$ denotes the nearest neighbour pair. $J> 0$ denotes the short-range interaction and $g$ sets the transverse field strength which fosters spin alignment along the $x$-axis. For time-independent case, we shall consider $J$ to be constant and equal to $1$ and $g=\frac{J}{2}$. Here also the uniformly random coefficients $h_{l}$ is chosen from a standard normal distribution on $[-1,1]$.

Similarly, the Hamiltonian for time-dependent TFQIM takes the following form
\begin{eqnarray}\label{tfqimtimedep}
    \mathcal{H}_\textsc{TFQIM}(t) &=& J(t)\sum_{l=1}^{n-1}Z_{l}Z_{l+1} + g(t)\sum_{l=1}^{n}X_{l}\\\nonumber && + \sum_{l=1}^{n}h_{l}Z_{l},
\end{eqnarray}
the value $J(t)$ oscillates between $1$ and $0$ depending on the even and odd time-steps and $g(t)=\frac{1}{2}$.

In Figure \ref{Heisenberg_Hamiltonian}, we illustrate the simulations for both the time-independent (equations (\ref{heis})) and time-dependent (equations (\ref{Heistimedep})) Heisenberg Hamiltonian $\mathcal{H}_{HEIS}$ respectively. In the time-independent scenario, we set $J=1$. The time-dependent parameter $g(t)$ is defined as $\sin{\frac{{\omega(t)}\pi}{2}}$, where $\omega(t)$ takes the values $1$ or $-1$ based on whether the time steps are even or odd during unitary evolution. Time is expressed in units of $\hbar$, and the time evolution spans $T=1.0$, with each time step set at $\frac{T}{r}=0.001$. Similarly, in Figure \ref{TFQIM_Hamiltonian}, we simulate both time-dependent and time-independent Hamiltonian $\mathcal{H}_{TFQIM}(t)$ as mentioned in equations (\ref{tfqim}) and (\ref{tfqimtimedep}) respectively. For the time-independent case, we shall consider $J$ to be constant and equal to $1$ and $g=\frac{J}{2}$. The time-dependent parameters $J(t)$ oscillate between $1$ and $0$ during even and odd time steps, and while $g(t)$ is set to be $\frac{1}{2}$. 

In both cases Figure \ref{Heisenberg_Hamiltonian} and Figure \ref{TFQIM_Hamiltonian}, the numerical Trotter error is illustrated using solid lines, each marked with different symbols to represent varying qubit counts --- 2 qubits (violet, $\circ$), 4 qubits (blue, $+$), 6 qubits (green, $\lhd$), 8 qubits (purple, $\diamond$), 10 qubits (red, $\times$), 12 qubits (teal-green, $\bigtriangleup$), and 14 qubits (orange, $\rhd$). As mentioned previously, the local Hamiltonian operators, being 1-sparse matrices, exhibit low-magnitude Frobenius norm errors, ranging between \(O(10^{-16})\) and \(O(10^{-12})\), refer Figure \ref{fig:fixeddepth}. The error associated with the proposed circuit method, represented by the black dotted line, closely follows the numerical error from the Trotter decomposition.

\begin{figure}[ht]
    \includegraphics[width=\linewidth]{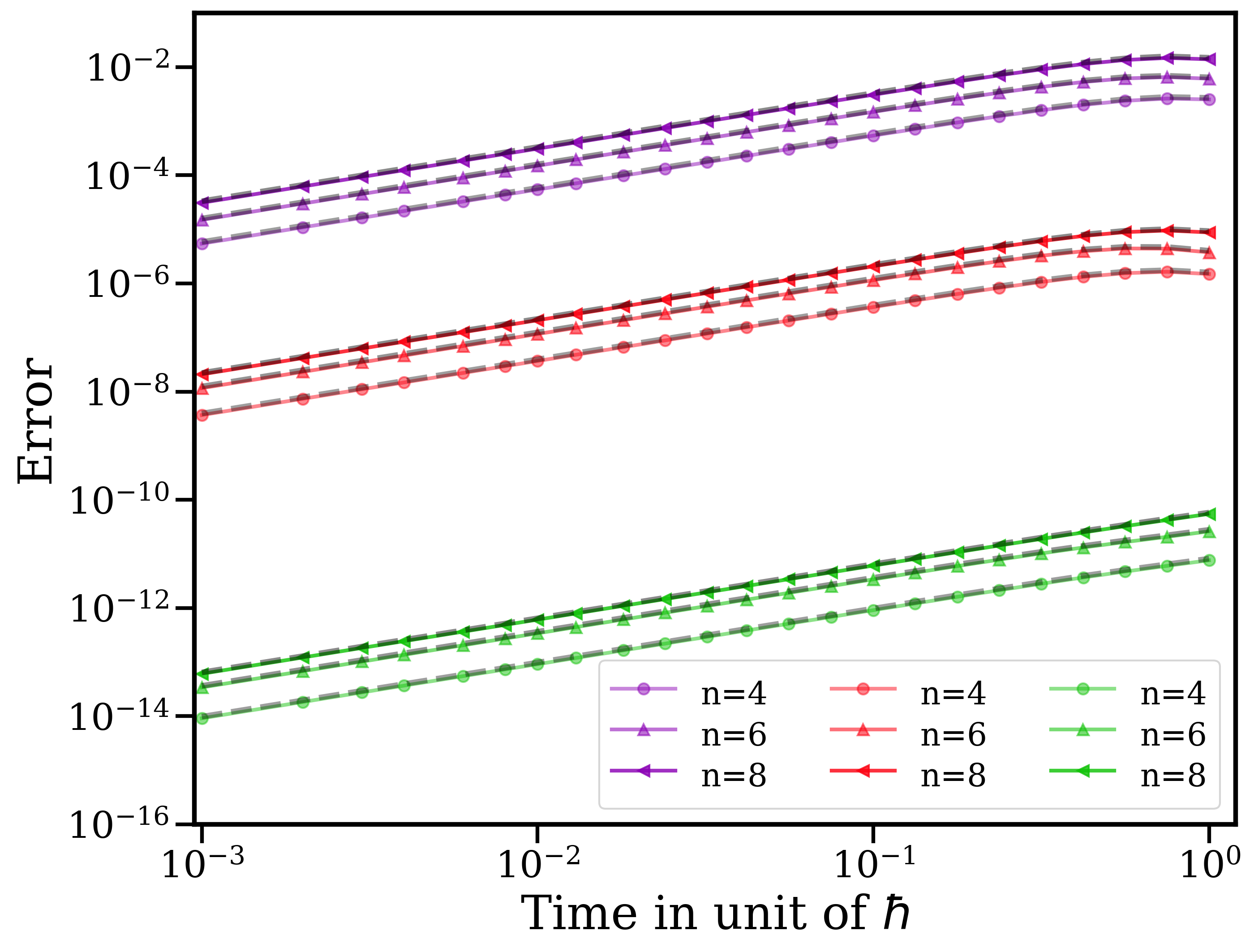}
    \caption{Comparison of the quantum circuit approximation error, represented by the Frobenius norm distance, for the unitary evolution of the time-independent transverse Magnetic Field Quantum Ising Hamiltonian using first, second and fourth order Trotter product formulas against their respective exact numerical errors. The plot displays numerical Trotter errors with solid lines, where distinct markers indicate different qubit counts, and colors represent Trotterization orders. The 4, 6, and 8 qubit systems are marked by \(\circ\), \(\bigtriangleup\), and \(\lhd\), respectively, while the first, second, and fourth-order Trotter formulas are shown in violet, red, and green colour. The black dotted line represents the error from the proposed circuit method, closely following the numerical Trotter error in all cases.}
    \label{higher_trotter}
\end{figure}

\subsection{Implementation of the proposed circuit method in higher order Trotter variants}

\begin{figure*}[ht]
    \includegraphics[width=0.95\textwidth]{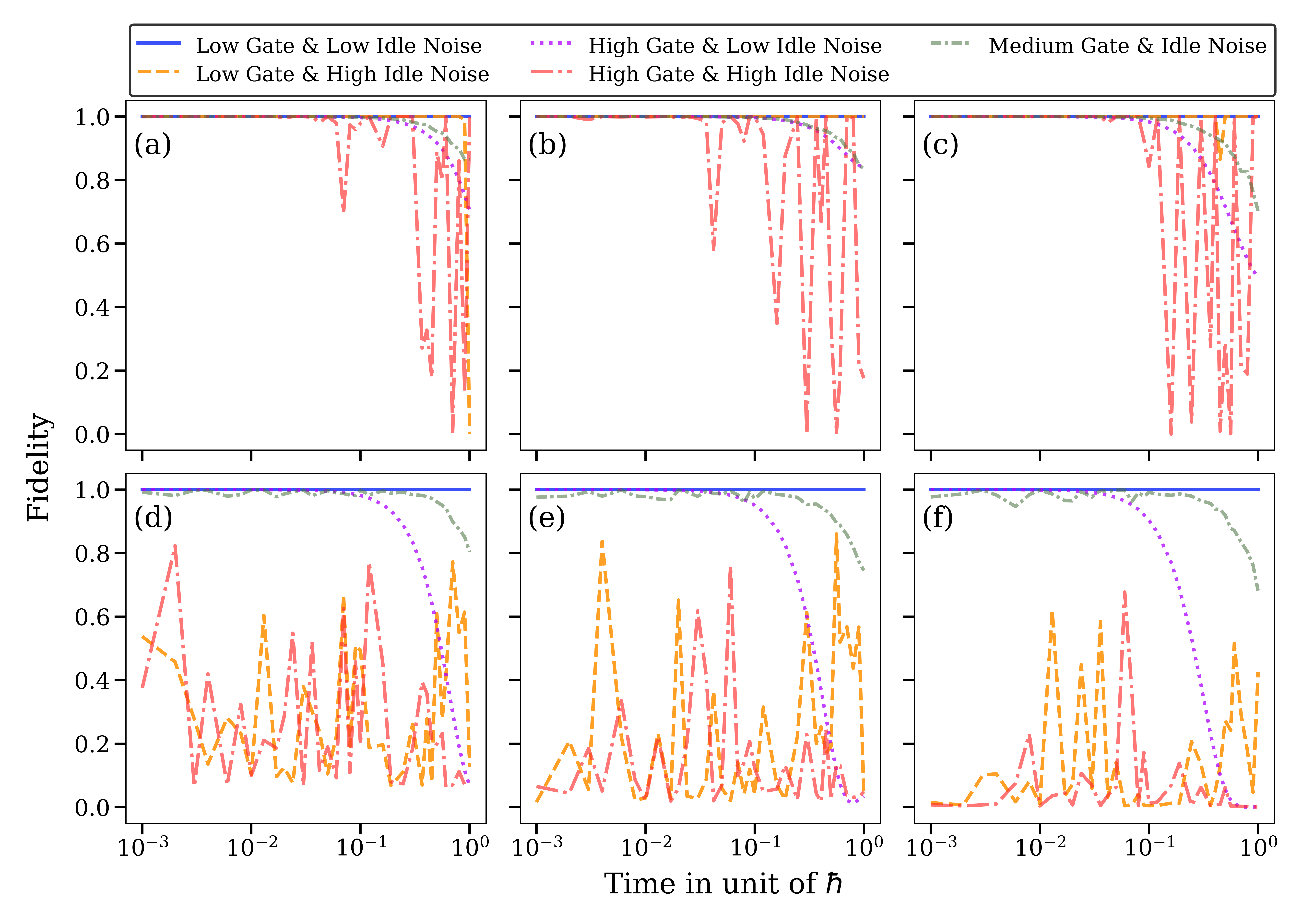}
  \caption{Fidelity $\mathcal{F}(t)$ between quantum states after passing through the noiseless and noisy circuit as a function of time and error strength for time-independent Random Field Heisenberg Hamiltonian. Figures (a), (b), (c) are with input state $\ket{\psi(0)}=\ket{1}^{\bigotimes n}$ for $4$, $6$ and $8$ qubits respectively, whereas Figures (d), (e), (f) are with input state $\ket{\psi(0)}=((\ket{0}+\ket{1})/\sqrt{2})^{\otimes n}$ for $4$, $6$ and $8$ qubits respectively. 
  }  \label{Fig:noisy1}
\end{figure*}

In this section, we further examine the proposed quantum circuit approximation method for unitary evolution operator $U(t)$ using higher-order viz. second and fourth order Trotter approximation formulae \cite{mc2023classically, kobayashi1994study}. Indeed, we restrict to report the simulation results for the time-independent transverse magnetic field quantum Ising Hamiltonian, the simulation results for other Hamiltonians considered in this paper are similar. It is important to note that although higher-order Trotter variants reduce the error, they also lead to an increase in the number of gates, a natural compromise. The details of higher-order Trotter is given in Appendix \ref{qc_hamil}.


The Figure \ref{higher_trotter} shows numerical Trotter errors as solid lines, with distinct markers indicating different qubit counts and colors representing Trotterization orders. The 4, 6, and 8 qubit systems are denoted by \(\circ\), \(\bigtriangleup\), and \(\lhd\), respectively, while the first, second, and fourth-order Trotter formulas are colored violet, red, and green. The black dotted line illustrates the error from the proposed circuit method, which closely mirrors the numerical Trotter error across all scenarios. The result from this section showcases that our proposed circuit method does not bring any additional error in the Hamiltonian simulation.

We like to emphasize that the total number of gates required for circuit simulations of a Hamiltonian is given by the number of terms in the Trooterization approximation formula times the number of gates to represent $\exp(\iota\theta\sigma)$ as described in Theorem \ref{gatecount}. For instance, a large value of $r$ in the first-order formula given by equation (\ref{trotter}) leads to huge number of gates in the circuit due to the  straight-forward implementation of Suzuki-Trotter product formula. However, identifying the clusters of commuting local Hamiltonians or Pauli-strings in the total Hamiltonian can significantly reduce the number of gates in use. A few techniques are introduced in literature to partitioning the set of all Pauli-strings into commutative and non-commutative clusters, see  \cite{Mukhopadhyay2023,Raeisi2012,vanden2020, Clinton2021, Zhong2022, Peng2022, Gokhale2023,Kurita2023,Boris2024}. However, these partitioning methods add up to higher computational complexities for complex Hamiltonians. We plan to explore the analysis on the construction of low-depth circuits by exploiting commuting Pauli-strings for complex Hailtonians in near future. 

\subsection{Noisy simulation}
\label{Noisy_simulation}

\begin{figure*}[ht]
    \includegraphics[width=0.95\textwidth]{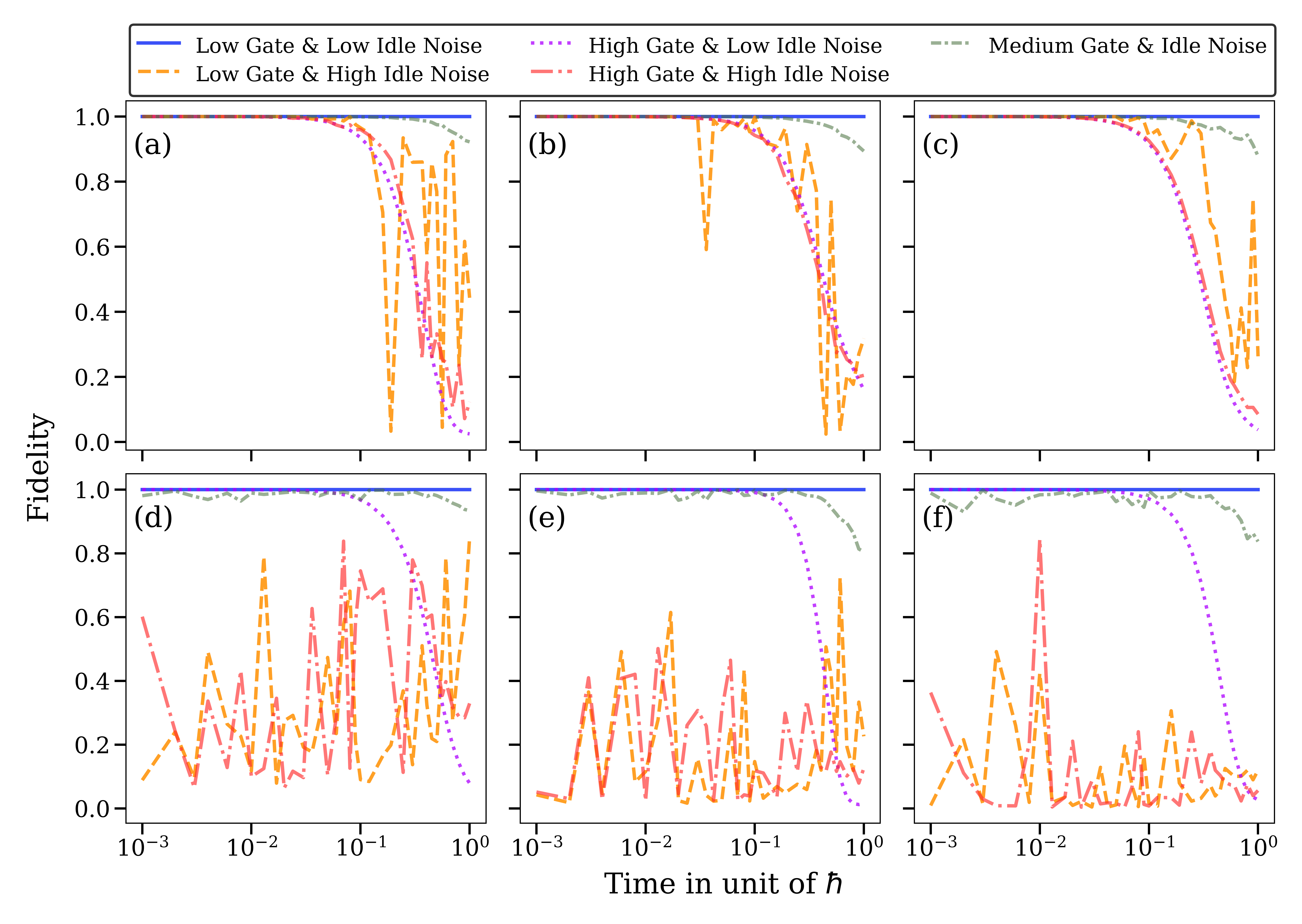}
  \caption{Fidelity $\mathcal{F}(t)$ between quantum states after passing through the noiseless and noisy circuit as a function of time and error strength for time-independent transverse magnetic random quantum Ising model  $\mathcal{H}_{TFQIM}$. Figures (a), (b), (c) are with input state $\ket{1}^{\bigotimes n}$ for $4$, $6$ and $8$ qubits respectively, whereas Figures (d), (e), (f) are with input state $((\ket{0}+\ket{1})/\sqrt{2})^{\otimes n}$ for $4$, $6$ and $8$ qubits respectively. 
  }  \label{Fig:noisy2}
\end{figure*}
 
A challenge in running simulations on current quantum hardware is controlling and preventing from random noise during implementation of quantum gates. To demonstrate the effectiveness of our methodologies in the presence of noise, in this section we incorporate different noise models in the circuit simulation of Hamiltonians that include two types of gate errors and two types of idle errors with varying strengths. The mathematical descriptions of these error channels are elaborated upon in Appendix \ref{Noise_Model_details}. The strength of each error channel, governed by the error probability $p$ in the descriptions of Kraus operators in Appendix \ref{Noise_Model_details}, is chosen at random from a uniform distribution. Furthermore, in our simulation, the strength of the noise is classified into three categories: low strength, with $p$ values ranging from $0$ to $10^{-3}$; medium strength, where $p$ lies between $10^{-3}$ to $10^{-1}$; and high strength, where the $p$ spans from $10^{-1}$ to $10^{0}$. 

To quantify the effect of noise, we consider the mean fidelity $\mathcal{F}(t)$, which measures the distance between the ideal (noise-free) and noisy output state vectors by calculating their squared overlap. Indeed, let $\ket{\psi(t)}$ and $\ket{{\psi(t)}}_{ns}$ denote the quantum states after time $t$ when $\ket{\psi(0)}$ is evolved using a noiseless and noisy circuit respectively. Then the fidelity \cite{bengtsson2017geometry} between $\ket{\psi(t)}$ and $\ket{{\psi}(t)}_{ns}$ is  defined as \begin{eqnarray}\label{fidelity}
    \mathcal{F}(t)={|\langle{\psi(t)}\ket{{\psi}(t)}_{ns}|^2}.
\end{eqnarray} 


Now using the Kraus operator formalism \cite{kraus1983states}, we perform the noisy circuit simulation with the initial states $(H\ket{0})^{\otimes n} =((\ket{0}+\ket{1})/\sqrt{2})^{\otimes n}$ and  $\ket{1}^{\bigotimes n}$. Figure \ref{Fig:noisy1}, \ref{Fig:noisy2} depict the results of these simulations under various scenarios: 1. Low gate and idle error strengths, 2. Low gate error but high idle error strengths (and vice versa), 3. High gate and idle error strengths, and 4. Medium gate and idle error strengths. In this context, `low,' `medium,' and `high' denote the error strengths. Gate errors encompass both bit-flip and phase-flip errors, while idle errors encompass both amplitude damping and phase damping.

In Figures \ref{Fig:noisy1}, \ref{Fig:noisy2} we observe the influence of the error of the gates in the quantum circuits  and the input state, while calculating the fidelity of the quantum state after passing through the noiseless and noisy quantum channel during time evolution. We note that when both gate and idle noise strength are low, the fidelity remains close to $1$ irrespective of the input state. Conversely, under other conditions---such as when either the gate or idle noise strength is medium, or when one is high while the other is low (or vice versa), or when both are high---we observe a deterioration in fidelity for the noisy quantum channel, particularly when the input state is $((\ket{0}+\ket{1})/\sqrt{2})^{\otimes n}$ compared to $\ket{1}^{\bigotimes n}$. Additionally, regardless of the input state, we find that when both gate and idle noise strengths are medium, the fidelity notably improves compared to the latter two scenarios where one (or both) of the gate and idle noise strengths are high. Only in Figure \ref{Fig:noisy1}, we observe that with the input state $\ket{1}^{\bigotimes n}$, the fidelity value $\mathcal{F}(t)$ for low gate and high idle noise strength surpasses the other two cases that are high gate and low(or high) idle noise strength. However, the fidelity value is deemed unacceptable or subpar for all other cases.

\section{Conclusion} \label{Conclusion}

In this paper, we proposed a scalable quantum circuit implementation of exponential of scaled Pauli-strings that are suitable for digital quantum computation. The circuits consist of one-qubit gates and $\cnot$ gates are designed in such a way that these are scalable and do not require the full connectivity for its implementation in the quantum hardware. We also justify that a quantum hardware with star-like quantum architecture can be used to efficiently implement the proposed circuits, which is obtained by showing the permutation similarity of certain Pauli-strings. 

Then these circuits are employed to simulate several time-independent and time-dependent 1D Hamiltonian operators by approximating its evolution using Suzuki-Trotter product formula. Indeed, it must be noted that the proposed methodology can be used for any other decomposition methods such as \textit{second-order symmetric Trotter formula} and \textit{order $2k$ symmetric Suzuki-Trotter formula}. Simulations involving up to $18$ qubits for $2$-sparse block-diagonal Hamiltonian, Quantum Ising model Hamiltonian and up to $14$ qubits for a class of other Hamiltonian operators, demonstrate that the error from the proposed circuit model simulation of Hamiltonian operators closely matches the numerical Trotter approximation error across all cases. Moreover, we observe that circuits of constant depth can handle the evolution of 1-sparse Hamiltonian and 2-sparse Ising Hamiltonian. 

Finally, we incorporate various noise models for gate errors and idle errors to assess the efficiency of the circuit models in NISQ computers for systems of up to $8$ qubits governed by time-independent Random Field Heisenberg Hamiltonian and time-independent transverse magnetic random quantum Ising models. These simulations show that the value of the fidelity $\mathcal{F}(t)$ between the outputs for noisy and noiseless simulation drops down to the $0.8$ mark during the end of the time evolution for certain states when the  magnitude of gate error and idle error ranges between $10^{-3}$ to $10^{-1}$ for different input states. Besides, when the evolution time is near to zero the value $1-\mathcal{F}(t) < 10^{-4}$ regardless of the input states. Finally, it is evident that the selection of the input state holds significant importance in noisy simulations. 

The proposed circuit models for exponential of Pauli-strings can be employed in diverse applications including simulation of higher dimensional Hamiltonian operators. The number of quantum gates for Hamiltonian simulation can be reduced by clubbing the commuting Pauli-strings into clusters before applying the Suzuki-Trotter formula. We plan to work on these problems in near future.\\





\section*{Acknowledgement} RSS and SC acknowledge support through the Prime Minister's Research Fellowship (PMRF), Government of India. The authors thank Prof. Sonjoy Majumder of IIT Kharagpur for inspiring discussions.




\bibliographystyle{unsrt}
\bibliography{references}

\begin{thebibliography}{10}

\bibitem{benioff1980computer}
Paul Benioff.
\newblock The computer as a physical system: A microscopic quantum mechanical hamiltonian model of computers as represented by turing machines.
\newblock {\em Journal of statistical physics}, 22:563--591, 1980.

\bibitem{manin1980computable}
Yuri Manin.
\newblock Computable and uncomputable.
\newblock {\em Sovetskoye Radio, Moscow}, 128:28, 1980.

\bibitem{feynman2018simulating}
Richard~P Feynman et~al.
\newblock Simulating physics with computers.
\newblock {\em Int. j. Theor. phys}, 21(6/7), 2018.

\bibitem{bharti2022noisy}
Kishor Bharti, Alba Cervera-Lierta, Thi~Ha Kyaw, Tobias Haug, Sumner Alperin-Lea, Abhinav Anand, Matthias Degroote, Hermanni Heimonen, Jakob~S Kottmann, Tim Menke, et~al.
\newblock Noisy intermediate-scale quantum algorithms.
\newblock {\em Reviews of Modern Physics}, 94(1):015004, 2022.

\bibitem{fauseweh2024quantum}
Benedikt Fauseweh.
\newblock Quantum many-body simulations on digital quantum computers: State-of-the-art and future challenges.
\newblock {\em Nature Communications}, 15(1):2123, 2024.

\bibitem{klein2006simulating}
Alexander Klein and Dieter Jaksch.
\newblock Simulating high-temperature superconductivity model hamiltonians with atoms in optical lattices.
\newblock {\em Physical Review A}, 73(5):053613, 2006.

\bibitem{hofstetter2018quantum}
Walter Hofstetter and Tao Qin.
\newblock Quantum simulation of strongly correlated condensed matter systems.
\newblock {\em Journal of Physics B: Atomic, Molecular and Optical Physics}, 51(8):082001, 2018.

\bibitem{peotta2021determination}
Sebastiano Peotta, Fredrik Brange, Aydin Deger, Teemu Ojanen, and Christian Flindt.
\newblock Determination of dynamical quantum phase transitions in strongly correlated many-body systems using loschmidt cumulants.
\newblock {\em Physical Review X}, 11(4):041018, 2021.

\bibitem{mcardle2020quantum}
Sam McArdle, Suguru Endo, Al{\'a}n Aspuru-Guzik, Simon~C Benjamin, and Xiao Yuan.
\newblock Quantum computational chemistry.
\newblock {\em Reviews of Modern Physics}, 92(1):015003, 2020.

\bibitem{lanyon2010towards}
Benjamin~P Lanyon, James~D Whitfield, Geoff~G Gillett, Michael~E Goggin, Marcelo~P Almeida, Ivan Kassal, Jacob~D Biamonte, Masoud Mohseni, Ben~J Powell, Marco Barbieri, et~al.
\newblock Towards quantum chemistry on a quantum computer.
\newblock {\em Nature chemistry}, 2(2):106--111, 2010.

\bibitem{poulin2011quantum}
David Poulin, Angie Qarry, Rolando Somma, and Frank Verstraete.
\newblock Quantum simulation of time-dependent hamiltonians and the convenient illusion of hilbert space.
\newblock {\em Physical review letters}, 106(17):170501, 2011.

\bibitem{low2018hamiltonian}
Guang~Hao Low and Nathan Wiebe.
\newblock Hamiltonian simulation in the interaction picture.
\newblock {\em arXiv preprint arXiv:1805.00675}, 2018.

\bibitem{low2017optimal}
Guang~Hao Low and Isaac~L Chuang.
\newblock Optimal hamiltonian simulation by quantum signal processing.
\newblock {\em Physical review letters}, 118(1):010501, 2017.

\bibitem{haah2021quantum}
Jeongwan Haah, Matthew~B Hastings, Robin Kothari, and Guang~Hao Low.
\newblock Quantum algorithm for simulating real time evolution of lattice hamiltonians.
\newblock {\em SIAM Journal on Computing}, (0):FOCS18--250, 2021.

\bibitem{Dong2022}
Yulong Dong, K.~Birgitta Whaley, and Lin Lin.
\newblock A quantum hamiltonian simulation benchmark.
\newblock {\em npj Quantum Information}, 8(1):131, 2022.

\bibitem{berry2015simulating}
Dominic~W Berry, Andrew~M Childs, Richard Cleve, Robin Kothari, and Rolando~D Somma.
\newblock Simulating hamiltonian dynamics with a truncated taylor series.
\newblock {\em Physical review letters}, 114(9):090502, 2015.

\bibitem{Zhong2022}
Jonathan Wei~Zhong Lau, Tobias Haug, Leong~Chuan Kwek, and Kishor Bharti.
\newblock {NISQ Algorithm for Hamiltonian simulation via truncated Taylor series}.
\newblock {\em SciPost Physics}, 12:122, 2022.

\bibitem{kieferova2019simulating}
M{\'a}ria Kieferov{\'a}, Artur Scherer, and Dominic~W Berry.
\newblock Simulating the dynamics of time-dependent hamiltonians with a truncated dyson series.
\newblock {\em Physical Review A}, 99(4):042314, 2019.

\bibitem{childs2019faster}
Andrew~M Childs, Aaron Ostrander, and Yuan Su.
\newblock Faster quantum simulation by randomization.
\newblock {\em Quantum}, 3:182, 2019.

\bibitem{faehrmann2022randomizing}
Paul~K Faehrmann, Mark Steudtner, Richard Kueng, Maria Kieferova, and Jens Eisert.
\newblock Randomizing multi-product formulas for hamiltonian simulation.
\newblock {\em Quantum}, 6:806, 2022.

\bibitem{sun2022perturbative}
Jinzhao Sun, Suguru Endo, Huiping Lin, Patrick Hayden, Vlatko Vedral, and Xiao Yuan.
\newblock Perturbative quantum simulation.
\newblock {\em Physical Review Letters}, 129(12):120505, 2022.

\bibitem{yuan2019theory}
Xiao Yuan, Suguru Endo, Qi~Zhao, Ying Li, and Simon~C Benjamin.
\newblock Theory of variational quantum simulation.
\newblock {\em Quantum}, 3:191, 2019.

\bibitem{yao2021adaptive}
Yong-Xin Yao, Niladri Gomes, Feng Zhang, Cai-Zhuang Wang, Kai-Ming Ho, Thomas Iadecola, and Peter~P Orth.
\newblock Adaptive variational quantum dynamics simulations.
\newblock {\em PRX Quantum}, 2(3):030307, 2021.

\bibitem{benedetti2021hardware}
Marcello Benedetti, Mattia Fiorentini, and Michael Lubasch.
\newblock Hardware-efficient variational quantum algorithms for time evolution.
\newblock {\em Physical Review Research}, 3(3):033083, 2021.

\bibitem{wang2021noise}
Samson Wang, Enrico Fontana, Marco Cerezo, Kunal Sharma, Akira Sone, Lukasz Cincio, and Patrick~J Coles.
\newblock Noise-induced barren plateaus in variational quantum algorithms.
\newblock {\em Nature communications}, 12(1):6961, 2021.

\bibitem{stilck2021limitations}
Daniel Stilck~Fran{\c{c}}a and Raul Garcia-Patron.
\newblock Limitations of optimization algorithms on noisy quantum devices.
\newblock {\em Nature Physics}, 17(11):1221--1227, 2021.

\bibitem{mc2023classically}
Conor Mc~Keever and Michael Lubasch.
\newblock Classically optimized hamiltonian simulation.
\newblock {\em Physical Review Research}, 5(2):023146, 2023.

\bibitem{Lotshaw2023}
Phillip~C. Lotshaw, Hanjing Xu, Bilal Khalid, Gilles Buchs, Travis~S. Humble, and Arnab Banerjee.
\newblock Simulations of frustrated ising hamiltonians using quantum approximate optimization.
\newblock {\em Philosophical Transactions of the Royal Society A: Mathematical, Physical and Engineering Sciences}, 381(2241):20210414, 2023.

\bibitem{PhysRevX.11.011020}
Andrew~M. Childs, Yuan Su, Minh~C. Tran, Nathan Wiebe, and Shuchen Zhu.
\newblock Theory of trotter error with commutator scaling.
\newblock {\em Phys. Rev. X}, 11:011020, Feb 2021.

\bibitem{childs2019nearly}
Andrew~M Childs and Yuan Su.
\newblock Nearly optimal lattice simulation by product formulas.
\newblock {\em Physical review letters}, 123(5):050503, 2019.

\bibitem{childs2018toward}
Andrew~M Childs, Dmitri Maslov, Yunseong Nam, Neil~J Ross, and Yuan Su.
\newblock Toward the first quantum simulation with quantum speedup.
\newblock {\em Proceedings of the National Academy of Sciences}, 115(38):9456--9461, 2018.

\bibitem{wiebe2014hamiltonian}
Nathan Wiebe, Christopher Granade, Christopher Ferrie, and David~G Cory.
\newblock Hamiltonian learning and certification using quantum resources.
\newblock {\em Physical review letters}, 112(19):190501, 2014.

\bibitem{shi2022parameterized}
Jinjing Shi, Wenxuan Wang, Xiaoping Lou, Shichao Zhang, and Xuelong Li.
\newblock Parameterized hamiltonian learning with quantum circuit.
\newblock {\em IEEE Transactions on Pattern Analysis and Machine Intelligence}, 45(5):6086--6095, 2022.

\bibitem{SarmaSarkar2023}
Rohit~Sarma Sarkar and Bibhas Adhikari.
\newblock Scalable quantum circuits for n-qubit unitary matrices.
\newblock In {\em 2023 IEEE International Conference on Quantum Computing and Engineering (QCE)}, volume~01, pages 1078--1088, 2023.

\bibitem{li2019tackling}
Gushu Li, Yufei Ding, and Yuan Xie.
\newblock Tackling the qubit mapping problem for nisq-era quantum devices.
\newblock In {\em Proceedings of the Twenty-Fourth International Conference on Architectural Support for Programming Languages and Operating Systems}, pages 1001--1014, 2019.

\bibitem{Raeisi2012}
Sadegh Raeisi, Nathan Wiebe, and Barry~C Sanders.
\newblock Quantum-circuit design for efficient simulations of many-body quantum dynamics.
\newblock {\em New Journal of Physics}, 14:103017, 2023.

\bibitem{vanden2020}
Ewout van~den Berg and Kristan Temme.
\newblock Circuit optimization of {H}amiltonian simulation by simultaneous diagonalization of {P}auli clusters.
\newblock {\em Quantum}, 4:322, 2020.

\bibitem{Clinton2021}
Laura Clinton, Johannes Bausch, and Toby Cubitt.
\newblock Hamiltonian simulation algorithms for near-term quantum hardware.
\newblock {\em Nature Communications}, 12(1):4989, 2021.

\bibitem{Mukhopadhyay2023}
Priyanka Mukhopadhyay, Nathan Wiebe, and Hong~Tao Zhang.
\newblock Synthesizing efficient circuits for hamiltonian simulation.
\newblock {\em npj Quantum Information}, 9, 2023.

\bibitem{Peng2022}
Bo~Peng, Sahil Gulania, Yuri Alexeev, and Niranjan Govind.
\newblock Quantum time dynamics employing the yang-baxter equation for circuit compression.
\newblock {\em Phys. Rev. A}, 106:012412, Jul 2022.

\bibitem{Yordanov2020}
Yordan~S. Yordanov, David R.~M. Arvidsson-Shukur, and Crispin H.~W. Barnes.
\newblock Efficient quantum circuits for quantum computational chemistry.
\newblock {\em Phys. Rev. A}, 102:062612, Dec 2020.

\bibitem{Reiher2017}
Markus Reiher, Nathan Wiebe, Krysta~M. Svore, Dave Wecker, and Matthias Troyer.
\newblock Elucidating reaction mechanisms on quantum computers.
\newblock {\em Proceedings of the National Academy of Sciences}, 114(29):7555--7560, 2017.

\bibitem{Kawase2022}
Yoshiaki Kawase and Keisuke Fujii.
\newblock Fast classical simulation of hamiltonian dynamics by simultaneous diagonalization using clifford transformation with parallel computation.
\newblock {\em Computer Physics Communications}, 288:108720, 2023.

\bibitem{Mansky2023}
M.~Mansky, V.~Puigvert, Castillo S., and C.~Linnhoff-Popien.
\newblock Decomposition algorithm of an arbitrary pauli exponential through a quantum circuit.
\newblock In {\em 2023 IEEE International Conference on Quantum Computing and Engineering (QCE)}, pages 434--442, Los Alamitos, CA, USA, 2023. IEEE Computer Society.

\bibitem{Ibm2017}
IBM~Q experience~backend information.
\newblock https://github.com/dcmckayibm/ibmqx-backend-information, 2017.

\bibitem{pires2021theoretical}
Antonio Sergio~Teixeira Pires.
\newblock {\em Theoretical tools for spin models in magnetic systems}.
\newblock IOP Publishing, 2021.

\bibitem{sarkar2024quantum}
Rohit~Sarma Sarkar and Bibhas Adhikari.
\newblock A quantum neural network framework for scalable quantum circuit approximation of unitary matrices.
\newblock {\em arXiv preprint arXiv:2405.00012}, 2024.

\bibitem{Qiskit}
{Qiskit contributors}.
\newblock Qiskit: An open-source framework for quantum computing, 2023,10.5281/zenodo.2573505.

\bibitem{Paramshakti}
PARAM Shakti High Performance~Computing Facility.
\newblock http://www.hpc.iitkgp.ac.in/hpcf/paramshakti.

\bibitem{ibmCircuitQuantum}
{I}{B}{M} {Q}uantum.
\newblock Quantum circuits | {I}{B}{M} {Q}uantum {D}ocumentation --- docs.quantum.ibm.com.
\newblock \url{https://docs.quantum.ibm.com/api/qiskit/circuit}.
\newblock [Accessed 28-02-2024].

\bibitem{kay2016quantum}
Alastair Kay.
\newblock Quantum error correction for state transfer in noisy spin chains.
\newblock {\em Physical Review A}, 93(4):042320, 2016.

\bibitem{backens2019jordan}
Stefan Backens, Alexander Shnirman, and Yuriy Makhlin.
\newblock Jordan--wigner transformations for tree structures.
\newblock {\em Scientific reports}, 9(1):2598, 2019.

\bibitem{fazekas1999lecture}
Patrik Fazekas.
\newblock {\em Lecture notes on electron correlation and magnetism}, volume~5.
\newblock World scientific, 1999.

\bibitem{skomski2008simple}
Ralph Skomski.
\newblock {\em Simple models of magnetism}.
\newblock Oxford university press, 2008.

\bibitem{kobayashi1994study}
Hiroto Kobayashi, Naomichi Hatano, and Masuo Suzuki.
\newblock Study of correction terms for higher-order decompositions of exponential operators.
\newblock {\em Physica A: Statistical Mechanics and its Applications}, 211(2-3):234--254, 1994.

\bibitem{Gokhale2023}
Pranav Gokhale, Olivia Angiuli, Yongshan Ding, Kaiwen Gui, Teague Tomesh, Martin Suchara, Margaret Martonosi, and Frederic~T Chong.
\newblock Minimizing state preparations in variational quantum eigensolver by partitioning into commuting families.
\newblock {\em arXiv preprint arXiv:1907.13623}, 2023.

\bibitem{Kurita2023}
Tomochika Kurita, Mikio Morita, Hirotaka Oshima, and Shintaro Sato.
\newblock Pauli string partitioning algorithm with the ising model for simultaneous measurements.
\newblock {\em The Journal of Physical Chemistry A}, 127(4):1068–1080, 2023.

\bibitem{Boris2024}
Boris Arseniev, Dmitry Guskov, Richik Sengupta, Jacob Biamonte, and Igor Zacharov.
\newblock Tridiagonal matrix decomposition for hamiltonian simulation on a quantum computer.
\newblock {\em Physical Review A}, 109(5):052627, 2024.

\bibitem{bengtsson2017geometry}
Ingemar Bengtsson and Karol {\.Z}yczkowski.
\newblock {\em Geometry of quantum states: an introduction to quantum entanglement}.
\newblock Cambridge university press, 2017.

\bibitem{kraus1983states}
Karl Kraus, Arno B{\"o}hm, John~D Dollard, and WH~Wootters.
\newblock {\em States, Effects, and Operations Fundamental Notions of Quantum Theory: Lectures in Mathematical Physics at the University of Texas at Austin}.
\newblock Springer, 1983.

\bibitem{hatano2005quantum}
Naomichi Hatano and Masuo Suzuki.
\newblock Finding exponential product formulas of higher orders.
\newblock {\em Quantum annealing and other optimization methods}, pages 37--68, 2005.

\bibitem{Ostmeyer2023optimized}
Johann Ostmeyer.
\newblock Optimised trotter decompositions for classical and quantum computing.
\newblock {\em Journal of Physics A: Mathematical and Theoretical}, 56(28):285303, 2023.

\bibitem{campbell2019random}
Earl Campbell.
\newblock Random compiler for fast hamiltonian simulation.
\newblock {\em Physical review letters}, 123(7):070503, 2019.

\bibitem{trotter1959product}
Hale~F Trotter.
\newblock On the product of semi-groups of operators.
\newblock {\em Proceedings of the American Mathematical Society}, 10(4):545--551, 1959.

\bibitem{rivas2012open}
Angel Rivas and Susana~F Huelga.
\newblock {\em Open quantum systems}, volume~10.
\newblock Springer, 2012.

\bibitem{nielsen2010quantum}
Michael~A Nielsen and Isaac~L Chuang.
\newblock {\em Quantum computation and quantum information}.
\newblock Cambridge university press, 2010.

\end{thebibliography}

\appendix
\begin{widetext}

\section{Understanding the permutation matrices in equation (\ref{permus1}) via an example} \label{appdx:A2}
The permutation matrices discussed in section \ref{ckt_model_Pauli_exp} comprise a certain sequence of $\cnot$ gates which follows Theorem \ref{SRBBvPauli}. To understand the CNOT connections for any Pauli string, let us take a specific example of $4$ qubit system. For a given Pauli string $XXII$, the binary representation of the string is $1100$. Thus, $b=12$. Thus $b/2=6$ with binary representation $110$. In such cases if $(x_2,x_1,x_0)=(110)$, this implies $m=\mbox{max}\{j | \delta_{1,x_j}=1, j=0,1,2\}=2$. Thus, following from our definition given in equation (\ref{permus1}), we obtain the following.

\begin{eqnarray}
    \Pi \mathsf{T}^{o}_{4,6}&=& (\cnot_{(1,4)}) (\cnot_{(4,2)}) \\
    &&(\cnot_{(4,1)}) (\cnot_{(1,4)}) \nonumber
\end{eqnarray}\normalsize i.e.\\
\centering
\Qcircuit @C=1em @R=.7em {&\lstick{1}&\ctrl{3}&\targ &\qw&\ctrl{3}&\qw\\
    &\lstick{2}&\qw&\qw &\targ&\qw&\qw\\&\lstick{3}&\qw&\qw &\qw&\qw&\qw\\  &\lstick{4}&\targ&\ctrl{-3} &\ctrl{-2}&\targ&\qw\\}
    
\end{widetext}

\begin{widetext}
\section{Proof of Theorem \ref{SRBBvPauli}}\label{appdx:A1}
\pf Let $\ket{\psi}=\ket{\psi_{1}\hdots \psi_n},$ $\ket{\psi_j}=a_j\ket{0}+b_j\ket{1}$ be an $n$-qubit input quantum state for unitary matrix represented by the product of sequence of quantum gates given by  $\Pi\mathsf{T}_{n,x}^g (I_2^{\otimes (n-1)}\otimes X) \Pi\mathsf{T}_{n,x}^g.$ Then we need to show that the output becomes $\sigma_b\ket{\psi}$ for some $b=(b_{1}\hdots b_n)\in\{0,1\}^n$ and $\sigma_b=\sigma_{1}\otimes\hdots\otimes\sigma_n$ with $\sigma_j=X$ if $b_k=1$ and $\sigma_k=I_2$ if $b_k=0.$


Let $x\in\{0,1,\hdots,2^{n-1}-1\}$ with the binary representation $x\equiv (x_{n-2},\hdots,x_{0})\in\{0,1\}^{n-1}.$ Suppose $J=\{j_1,\hdots,j_l\}\subseteq \{0,\hdots,n-2\}$ such that $x_j=1$ if $j\in J$ and $x_j=0$ if $j\notin J.$ Further, let $j_1<j_2\hdots<j_l$. Now first consider $g=e.$ Then


\begin{eqnarray*}
&& \Pi\mathsf{T}_{n,x}^e (I_2^{\otimes n-1}\otimes X) \Pi\mathsf{T}_{n,x}^e \ket{\psi} \\ 
&=& \Pi\mathsf{T}_{n,x}^e (I_2^{\otimes n-1}\otimes X) \Pi\mathsf{T}_{n,x}^e (a_n\ket{\psi_{1}\hdots\psi_{n-1}0} + b_n\ket{\psi_{1}\hdots\psi_{n-1}1}) \\
&=& \Pi\mathsf{T}_{n,x}^e (I_2^{\otimes n-1}\otimes X)\left(a_n\ket{\psi_{1}\hdots\psi_{n-1}0} + b_n\ket{\psi_{1}\hdots (X\psi_{n-j_l-1})\hdots (X\psi_{n-j_1-1})\hdots 1}\right) \\
&=& \Pi\mathsf{T}_{n,x}^e \left(a_n\ket{\psi_{1}\hdots\psi_{n-1}1} + b_n\ket{\psi_{1}\hdots (X\psi_{n-j_l-1})\hdots (X\psi_{n-j_1-1})\hdots 0}\right) \\
&=& a_n\ket{\psi_{1}\hdots (X\psi_{n-j_l-1})\hdots (X\psi_{n-j_1-1})\hdots 1} + b_0\ket{\psi_{n-1}\hdots (X\psi_{n-j_l-1})\hdots (X\psi_{n-j_1-1})\hdots 0} \\
&=& \ket{\psi_{1}\hdots (X\psi_{n-j_l-1})\hdots (X\psi_{n-j_1-1})\hdots (X\psi_n)} \\
&=&\sigma_b\ket{\psi},
\end{eqnarray*} where $\sigma_b=\sigma_{1}\otimes \hdots\otimes\sigma_n$ such that $\sigma_{n-j-1}=X$ if $j\in J$ and $I_2$ otherwise and also $\sigma_n=X$. From our definition $x=\sum_{j\in J}2^j$ and  $b=\sum_{j\in J}2^{n-(n-j-1)}+1=\sum_{j\in J}2^{j+1}+1$ is odd and thus $x=\frac{b-1}{2}$.
Now taking $g=o$,
\begin{eqnarray*}
&& \Pi\mathsf{T}_{n,x}^o (I_2^{\otimes n-1}\otimes X) \Pi\mathsf{T}_{n,x}^o \ket{\psi} \\ 
&=& (\cnot)_{(n-j_l-1,n)}\Pi\mathsf{T}_{n,x}^e (\cnot)_{(n-j_l-1,n)}(I_2^{\otimes n-1}\otimes X)\\&& (\cnot)_{(n-j_l-1,n)}\Pi\mathsf{T}_{n,x}^e(\cnot)_{(n-j_l-1,n)} \ket{\psi}\\%
&=& (\cnot)_{(n-j_l-1,n)}\Pi\mathsf{T}_{n,x}^e (\cnot)_{(n-j_l-1,n)}(I_2^{\otimes n-1}\otimes X) \\&&(\cnot)_{(n-j_l-1,n)}\Pi\mathsf{T}_{n,x}^e(\cnot)_{(n-j_l-1,n)}\\&&(a_{n-j_l-1}\ket{\psi_{1}\hdots\psi_{n-j_l-2}{0}\psi_{n-j_l}\hdots\psi_{n}} + b_{n-j_l-1}\ket{\psi_{1}\hdots\psi_{n-j_l-2}{1}\psi_{n-j_l}\hdots\psi_{n}}) \\
&=& (\cnot)_{(n-j_l-1,n)}\Pi\mathsf{T}_{n,x}^e (\cnot)_{(n-j_l-1,n)}(I_2^{\otimes n-1}\otimes X) (\cnot)_{(n-j_l-1,n)}\Pi\mathsf{T}_{n,x}^e\\ &&(a_{n-j_l-1}\ket{\psi_{1}\hdots\psi_{n-j_l-2}{0}\psi_{n-j_l}\hdots\psi_{n}} + b_{n-j_l-1}\ket{\psi_{1}\hdots\psi_{n-j_l-2}{1}\psi_{n-j_l}\hdots(X\psi_{n}})) \\
&=& (\cnot)_{(n-j_l-1,n)}\Pi\mathsf{T}_{n,x}^e (\cnot)_{(n-j_l-1,n)}(I_2^{\otimes n-1}\otimes X) (\cnot)_{(n-j_l-1,n)}\Pi\mathsf{T}_{n,x}^e\\ &&(a_{n-j_l-1}a_n\ket{\psi_{1}\hdots\psi_{n-j_l-2}{0}\psi_{n-j_l}\hdots 0} + a_{n-j_l-1}b_n\ket{\psi_{1}\hdots\psi_{n-j_l-2}{0}\psi_{n-j_l}\hdots 1}\\&&+b_{n-j_l-1}a_n\ket{\psi_{1}\hdots\psi_{n-j_l-2}{1}\psi_{n-j_l}\hdots {0}} +b_{n-j_l-1}b_n\ket{\psi_{1}\hdots\psi_{n-j_l-2}{1}\psi_{n-j_l}\hdots {1}}) \\
&=& (\cnot)_{(n-j_l-1,n)}\Pi\mathsf{T}_{n,x}^e (\cnot)_{(n-j_l-1,n)}(I_2^{\otimes n-1}\otimes X) (\cnot)_{(n-j_l-1,n)}\\ &&(a_{n-j_l-1}a_n\ket{\psi_{1}\hdots\psi_{n-j_l-2}{0}\psi_{n-j_l}\hdots 0} \\&&+ a_{n-j_l-1}b_n\ket{\psi_{1}\hdots\psi_{n-j_l-2}{1}\psi_{n-j_l}\hdots (X\psi_{n-j_1-1})\hdots 1}\\&&+b_{n-j_l-1}a_n\ket{\psi_{1}\hdots\psi_{n-j_l-2}{1}\psi_{n-j_l}\hdots {0}} \\&&+b_{n-j_l-1}b_n\ket{\psi_{1}\hdots\psi_{n-j_l-2}{0}\psi_{n-j_l}\hdots (X\psi_{n-j_1-1})\hdots {1}}) \\
&=& (\cnot)_{(n-j_l-1,n)}\Pi\mathsf{T}_{n,x}^e (\cnot)_{(n-j_l-1,n)}(I_2^{\otimes n-1}\otimes X) \\ &&(a_{n-j_l-1}a_n\ket{\psi_{1}\hdots\psi_{n-j_l-2}{0}\psi_{n-j_l}\hdots 0} \\&&+ a_{n-j_l-1}b_n\ket{\psi_{1}\hdots\psi_{n-j_l-2}{1}\psi_{n-j_l}\hdots (X\psi_{n-j_1-1})\hdots 0}\\&&+b_{n-j_l-1}a_n\ket{\psi_{1}\hdots\psi_{n-j_l-2}{1}\psi_{n-j_l}\hdots {1}} \\&&+b_{n-j_l-1}b_n\ket{\psi_{1}\hdots\psi_{n-j_l-2}{0}\psi_{n-j_l}\hdots (X\psi_{n-j_1-1})\hdots {1}}) \\
&=& (\cnot)_{(n-j_l-1,n)}\Pi\mathsf{T}_{n,x}^e (\cnot)_{(n-j_l-1,n)}\\ &&(a_{n-j_l-1}a_n\ket{\psi_{1}\hdots\psi_{n-j_l-2}{0}\psi_{n-j_l}\hdots 1} \\&&+ a_{n-j_l-1}b_n\ket{\psi_{1}\hdots\psi_{n-j_l-2}{1}\psi_{n-j_l}\hdots (X\psi_{n-j_1-1})\hdots 1}\\&&+b_{n-j_l-1}a_n\ket{\psi_{1}\hdots\psi_{n-j_l-2}{1}\psi_{n-j_l}\hdots {0}} \\&&+b_{n-j_l-1}b_n\ket{\psi_{1}\hdots\psi_{n-j_l-2}{0}\psi_{n-j_l}\hdots (X\psi_{n-j_1-1})\hdots {0}}) \\
&=& (\cnot)_{(n-j_l-1,n)}\Pi\mathsf{T}_{n,x}^e \\ &&(a_{n-j_l-1}a_n\ket{\psi_{1}\hdots\psi_{n-j_l-2}{0}\psi_{n-j_l}\hdots 1} \\&&+ a_{n-j_l-1}b_n\ket{\psi_{1}\hdots\psi_{n-j_l-2}{1}\psi_{n-j_l}\hdots (X\psi_{n-j_1-1})\hdots 0}\\&&+b_{n-j_l-1}a_n\ket{\psi_{1}\hdots\psi_{n-j_l-2}{1}\psi_{n-j_l}\hdots {1}} \\&&+b_{n-j_l-1}b_n\ket{\psi_{1}\hdots\psi_{n-j_l-2}{0}\psi_{n-j_l}\hdots (X\psi_{n-j_1-1})\hdots {0}}) \\
    &=& (\cnot)_{(n-j_l-1,n)}(a_{n-j_l-1}a_n\ket{\psi_{1}\hdots\psi_{n-j_l-2}{1}\psi_{n-j_l}\hdots (X\psi_{n-j_1-1})\hdots 1}\\&& + a_{n-j_l-1}b_n\ket{\psi_{1}\hdots\psi_{n-j_l-2}{1}\psi_{n-j_l}\hdots (X\psi_{n-j_1-1})\hdots 0} \\&&+b_{n-j_l-1}a_n\ket{\psi_{1}\hdots\psi_{n-j_l-2}{0}\psi_{n-j_l}\hdots (X\psi_{n-j_1-1})\hdots{1}}\\&&+b_{n-j_l-1}b_n\ket{\psi_{1}\hdots\psi_{n-j_l-2}{0}\psi_{n-j_l}\hdots (X\psi_{n-j_1-1})\hdots {0}}) \\
&=& (a_{n-j_l-1}a_n\ket{\psi_{1}\hdots\psi_{n-j_l-2}{1}\psi_{n-j_l}\hdots (X\psi_{n-j_1-1})\hdots 0}\\&& + a_{n-j_l-1}b_n\ket{\psi_{1}\hdots\psi_{n-j_l-2}{1}\psi_{n-j_l}\hdots (X\psi_{n-j_1-1})\hdots 1} \\&&+b_{n-j_l-1}a_n\ket{\psi_{1}\hdots\psi_{n-j_l-2}{0}\psi_{n-j_l}\hdots (X\psi_{n-j_1-1})\hdots{1}}\\&&+b_{n-j_l-1}b_n\ket{\psi_{1}\hdots\psi_{n-j_l-2}{0}\psi_{n-j_l}\hdots (X\psi_{n-j_1-1})\hdots {0}}) \\
&=& \ket{\psi_{1}\hdots (X\psi_{n-j_l-1})\hdots (X\psi_{n-j_1-1})\hdots \psi_n} =\sigma_b\ket{\psi},
\end{eqnarray*}
where $\sigma_b=\sigma_{1}\otimes \hdots\otimes\sigma_n$ such that $\sigma_{n-j-1}=X$ if $j\in J$ and $I_2$ otherwise. From our definition $x=\sum_{j\in J}2^j$ and $b=\sum_{j\in J}2^{n-(n-j-1)}=\sum_{j\in J}2^{j+1}$ is even and thus $x=\frac{b}{2}$. When $x=0$, the case is trivial and this completes the proof. $\hfill\square$
\end{widetext}

\begin{widetext}
\section{Suzuki-Trotter expansion via quantum circuits}\label{qc_hamil}

Hamiltonian simulation is performed by approximating the time-evolution unitary at each time step via a quantum circuit by employing a decomposition of the unitary such as Suzuki-Trotter decomposition scheme or Trotterization \cite{hatano2005quantum,Ostmeyer2023optimized}. The first-order Trotterization given by 
\begin{equation}\label{trotter}
    U(T)  = e^{-\iota HT}
       \approx \left( e^{-\iota{H}T/r} \right)^{r}
       \approx\left( \prod_{j=1}^{L}e^{-\iota {a_j}{H_{j}}T/r} \right)^{r}
\end{equation}
for a time-independent Hamiltonian given by $H=\sum_{j=1}^{L} a_j H_j$ with  $O(T^{2}/r^2)$ approximation error for some positive integer $r$, known as the Trotter parameter \cite{poulin2011quantum}. Thus, to simulate the Hamiltonian over a long time $T$, the evolution is segmented into $r$ Trotter steps, while preserving an error level of at most $\epsilon/r$, where $\epsilon>0$ such that the Trotter approximation error $\left\|U(T)-\left( \prod_{j=1}^{L}e^{-\iota{a_j}{H_{j}}T/r} \right)^{r}\right\|_F\leq \epsilon.$ However, in practice, it is crucial to decide the value of the Trotter parameter $r$ while performing the simulation.  It is known that the first-order deterministic Trotter error cannot exceed $O\left((T\lambda L)^{2}/\epsilon r^2\right)$, where $\lambda := \max_{j}\left\| H_{j} \right\|_2$, $L$ is the total number of local Hamiltonian operators \cite{childs2019faster} \cite{campbell2019random} \cite{PhysRevX.11.011020}. Now, according to Equation \ref{trotter}, quantum circuits generate a succession of blocks responsible for performing specific time evolution, with their depth increasing linearly with $r$.

For time-dependent Hamiltonians $H(t)=\sum_{j=1}^{L} a_j(t) H_j$ where $a_j(t)$ is non-constant for some values of $j$, then two approximations are employed to simulate the time evolution operator \cite{trotter1959product} \cite{poulin2011quantum}. It is first approximated as a piece-wise constant function by discretizing the time interval $[0,\, T]$ into small intervals of length $\Delta t$ during which the time-dependent Hamiltonian remains constant, and in the second, during that time interval, the matrix exponential of the Hamiltonian is approximated  using first-order Trotter decompositions as discussed above \cite{trotter1959product} \cite{poulin2011quantum}. Thus the time evolution operator is approximated as
\begin{equation}\label{eqn:tdh}
    U(T) \approx \prod_{k=1}^{r} \exp\left({-\iota H(k\Delta t)\Delta t}\right)
\end{equation}
where $k$ multiplies over the number of discretized time-steps $\Delta t = T/r$, $r$ is the number of time-steps of length $\Delta t$. The choice of $r$ needs to ensure that $\Delta t$ should be less than the fluctuation time-scale of $H(\Delta t)$ and the decomposition error is less than $\epsilon$. Indeed, for time-dependent cases $a_j(t)$ is a piece-wise constant function i.e. $a_j(t)=a_j(k\Delta t), t\in [(k-1)\Delta t,k\Delta t], k\in \{1,\hdots,r\}$. Thus $U(T)\approx \prod_{k=1}^r e^{-\iota H(k\Delta t)\Delta t}\approx \prod_{k=1}^r \prod_{j=1}^Le^{-\iota a_j(k\Delta t)H_j\Delta t }.$

\begin{figure*}[t]
    \centering
    \includegraphics[height=2cm,width=8.6cm]{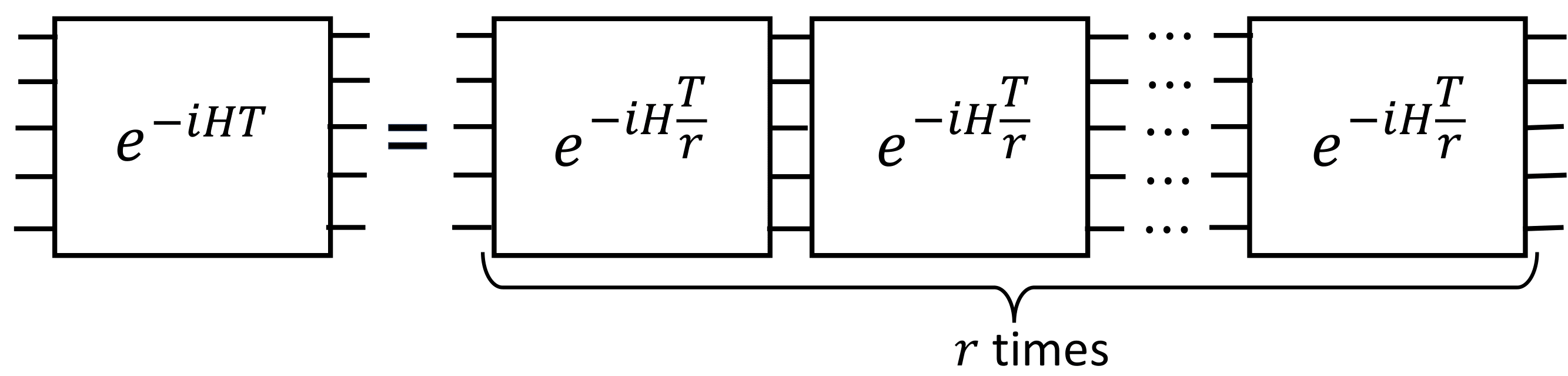}
    \caption{Trotter approximation through quantum circuit}
\end{figure*}

Indeed the higher-order Trotter generates more accurate decomposition of a Hamiltonian due to incurring a higher power of error expression. We recall that the ($2k$)th-order Suzuki-Trotter formula $U_{2k}(T) = e^{-iTH} + \mathcal{O}((T/ r)^{2k+1})$ is defined via recursion \cite{kobayashi1994study} as
\begin{eqnarray*}
    U_2(T) &:=& ({e^{-\frac{i T}{2r} a_1H_1} \hdots e^{-\frac{i T}{2r} {a_L H_{L}}}e^{-\frac{i T}{2r} {a_LH_{L}}} \hdots e^{-\frac{i T}{2r} a_1 H_1}})^r, \\
    U_{2k}(T) &:=& ({U_{2k-2}(\nu_k \frac{T}{r})}^2 U_{2k-2}((1 - 4\nu_k) \frac{T}{r}) {U_{2k-2}(\nu_k \frac{T}{r})}^2)^r
\end{eqnarray*}
where $\nu_k := \frac{1}{4 - 4^{1/(2k-1)}}$.

Thus, simulating a Hamiltonian through quantum circuits boils down to efficient quantum circuit implementation of evolution of local Hamiltonian operators given by $\exp(-\iota H_j\theta),$ for some constant $\theta> 0.$  Since the parametrized one-qubit gates and two-qubit control gates form a universal model of quantum computation, quantum circuits with optimal number of such gates for scalable implementation of exponential of scaled Pauli-strings, which are the local Hamiltonians in general, is the fundamental matter of concern here. Upon achieving this, the quantum circuit design employing equations (\ref{trotter}) and (\ref{eqn:tdh}), the probability distribution corresponding to the desired quantum state $\ket{\psi(T)}$ can be obtained for the input state $\ket{\psi(0)}$ after the quantum measurement of the output state of the quantum circuit with respect to computational basis.
    
\end{widetext}

\begin{widetext}
\section{Noise Model}
\label{Noise_Model_details}



Generally, the noise in a quantum circuit is considered of two types: per-gate noise and per-time noise \cite{Clinton2021}. In the first case, the noise is available at a constant rate per gate, independent of the time it takes to implement that gate which is a standard practice in quantum computation and the model which we have implemented in this paper. In the second noise model, it is considered that noise occurs at a constant rate per unit time and is mainly used in developing error models in physics which we consider in this paper.

Now, it is to be noted that several techniques have been developed to analyze the state evolution of a quantum system in the presence of noise \cite{rivas2012open}. Among these, the Kraus operator formalism \cite{kraus1983states} stands out as one of the most prevalent methods. Under this formalism the evolution of a quantum state through a noisy quantum circuit is given by \cite{nielsen2010quantum},
\begin{equation}
    \rho{'} = \sum_{j} E_{j} \rho E_{j}^{\dagger},
\label{Kraus_matrix}    
\end{equation}
where $\rho = \ket{\psi}\bra{\psi}$ i.e. the density matrix; as we are only dealing with pure quantum state $\ket{\psi}$ and the $E_{j}$ represent Kraus operators associated with various noise channels, satisfying the condition $\sum_{j} E_{j} E_{j}^{\dagger} = I$, where $I$ denotes the identity operator \cite{nielsen2010quantum}. The density matrices $\rho{'}$ and $\rho$ correspond to the states after and before undergoing a noisy quantum channel, respectively. In our study, we restrict ourselves to four noisy quantum channels described below, where $p$ is error probability such that $0 \leq p \leq 1.$

\begin{enumerate}

\item Gate error: The Kraus operators corresponding to the bit-flip channel are $E_0=\sqrt{1-p}I_2, E_1=\sqrt{p}X,$ and for the phase-flip channel $E_0=\sqrt{1-p}X, E_1=\sqrt{p}Z.$ 

\item Idle error: The Kraus operators corresponding to the amplitude-damping channel are $E_{0}=\begin{pmatrix}
1 & 0\\ 
0 & {\sqrt{1-p}}
\end{pmatrix},$ $E_{1}=\begin{pmatrix}
1 & {\sqrt{p}}\\ 
0 & 0
\end{pmatrix}$, and for the Phase damping channel $E_{0}=\begin{pmatrix}
1 & 0\\ 
0 & {\sqrt{1-p}}
\end{pmatrix},$ $E_{1}=\begin{pmatrix}
1 & 0\\ 
0 & {\sqrt{p}}
\end{pmatrix}.$
\end{enumerate}

We incorporate these error models in the proposed quantum circuit simulation with varying degrees of error in order to observe their impact on the circuit and in turn, simulation of the Hamiltonian, see Section \ref{Noisy_simulation}.
    
\end{widetext}

\end{document}